\pdfoutput=1
\documentclass[11pt,a4]{JHEP3}

\usepackage[utf8]{inputenc} 
\usepackage{cite} 
\usepackage{amsmath} 
\usepackage{amssymb} 
\usepackage{mathrsfs} 
\usepackage{graphicx} 
\usepackage{todonotes} 
\usepackage{xcolor} 
\usepackage{algpseudocode} 
\let\normalcolor\relax 
\usepackage{textgreek} 
\newcommand{\cF}{\mathcal{F}}

\newcommand{\cH}{\mathcal{H}}
\newcommand{\cA}{\mathcal{A}}

\newcommand{\cN}{\mathcal{N}}

\newcommand{\eq}[1]{\begin{equation}#1\end{equation}}
\newcommand{\eqs}[1]{\begin{equation}\begin{split}#1\end{split}\end{equation}}
\newcommand{\eref}[1]{equation~(\ref{#1})} 
\newcommand{\sref}[1]{section~\ref{#1}} 
\newcommand{\Sref}[1]{Section~\ref{#1}} 
\newcommand{\fref}[1]{Figure~\ref{#1}} 
\newcommand{\aref}[1]{appendix~\ref{#1}} 
\newcommand{\Aref}[1]{Appendix~\ref{#1}} 

\makeatletter
\newcommand*{\textoverline}[1]{$\overline{\hbox{#1}}\m@th$}
\makeatother
\newcommand{\MHVB}{\textoverline{MHV} {}} 

\newcommand{\diracd}[2]{\delta^{#1}\left(#2\right)}

\newcommand{\ang}[1]{\left\langle #1\right\rangle}
\newcommand{\squ}[1]{\left[ #1\right]}
\newcommand{\bra}[1]{\left\langle #1\right|}
\newcommand{\ket}[1]{\left| #1\right\rangle}
\newcommand{\sbra}[1]{\left[ #1\right|}
\newcommand{\sket}[1]{\left| #1\right]}

\newcommand{\pdd}[2]{\frac{\partial #1}{\partial #2}}

\definecolor{inoutcolor}{rgb}{0.65,0.65,0.65}
\definecolor{varcolor}{rgb}{0.1,0.35,0.75}
\definecolor{functioncolor}{rgb}{0.1,0.35,0.75}
\newcommand{\var}[1]{\textit{#1}}
\newcommand{\fun}[1]{{\tt\color{functioncolor}#1}}
\newcommand{\mathematica}[2]{
\vspace{0.1cm}
\noindent\boxed{\begin{minipage}{\textwidth}\begin{tabular}{lp{\textwidth-2cm}}
{\tt \color{inoutcolor}\scriptsize In[1]:= }&{\tt #1}\\
{\tt \color{inoutcolor}\scriptsize Out[1]= }&{\tt #2}
\end{tabular}\end{minipage}}
\vspace{0.3cm}
}
\newcommand{\mathematicaa}[4]{
\vspace{0.1cm}
\noindent\boxed{\begin{minipage}{\textwidth}\begin{tabular}{lp{\textwidth-2cm}}
{\tt \color{inoutcolor}\scriptsize In[1]:= }&{\tt #1}\\
{\tt \color{inoutcolor}\scriptsize Out[1]= }&{\tt #2}\\[.3cm]
{\tt \color{inoutcolor}\scriptsize In[2]:= }&{\tt #3}\\
{\tt \color{inoutcolor}\scriptsize Out[2]= }&{\tt #4}
\end{tabular}\end{minipage}}
\vspace{0.3cm}
}
\newcommand{\mathematicaaa}[6]{
\vspace{0.1cm}
\noindent\boxed{\begin{minipage}{\textwidth}\begin{tabular}{lp{\textwidth-2cm}}
{\tt \color{inoutcolor}\scriptsize In[1]:= }&{\tt #1}\\
{\tt \color{inoutcolor}\scriptsize Out[1]= }&{\tt #2}\\[.3cm]
{\tt \color{inoutcolor}\scriptsize In[2]:= }&{\tt #3}\\
{\tt \color{inoutcolor}\scriptsize Out[2]= }&{\tt #4}\\[.3cm]
{\tt \color{inoutcolor}\scriptsize In[3]:= }&{\tt #5}\\
{\tt \color{inoutcolor}\scriptsize Out[3]= }&{\tt #6}
\end{tabular}\end{minipage}}
\vspace{0.3cm}
}

\date{}
\title{\Large A Monte Carlo Approach to the 4D Scattering Equations}
\author{Joseph A. Farrow\vspace{7pt}\\\normalsize\textit{
Department of Mathematical Sciences}\\\normalsize \textit{\,\,Durham University, Durham, DH1 3LE, United Kingdom}}
\preprint{June 2018}
\abstract{The scattering equation formalism is a general framework for calculation of amplitudes in theories of massless particles. We provide a detailed introduction to the 4D scattering equation framework accessible to non-experts, outline current difficulties solving the equations numerically, and explain how to overcome them with a Monte Carlo algorithm. With this submission we include {\tt treeamps4dJAF}, the first publicly available {\sc Mathematica} package for calculating amplitudes by solving the scattering equations, supporting MHV analytical and N$^{k-2}$MHV numerical computations. The package provides a powerful and flexible computational tool for calculating tree-level amplitudes in super Yang Mills theories, Einstein supergravity and conformal supergravity.  We tabulate sets of numerical solutions up to 9 points in all MHV sectors and 12 points in the NHMV sector which can be used for fast evaluation of amplitudes.}

\begin{document}

\maketitle

\tableofcontents
\pagebreak

\section{Introduction}
The calculation of scattering amplitudes in quantum field theories is both an essential ingredient in interpreting data from high energy physics experiment, and a powerful tool for probing the deepest questions of theoretical physics. The development of new mathematical techniques for calculating amplitudes has lead to important advances in both of these areas. In the 1980s the introduction of spinor helicity notation gave way to new simplified computations of four dimensional amplitudes which previously had seemed intractable, for example \cite{Xu:1986xb, Kleiss:1985yh,Parke:1986gb}. In recent years there have been many advances in the set of technical tools that exist for calculating perturbative scattering amplitudes. Key tree-level techniques include recursive methods for calculation of higher point amplitudes from lower point inputs, known as BCFW recursion\cite{Britto:2004ap, Britto:2005fq}, and the formulation of field theory amplitudes in terms of string worldsheet calculations, known as twistor-string theory\cite{Witten:2003nn,Berkovits:2004hg,Roiban:2004yf}.

The simplicity of the Parke-Taylor form of the Yang-Mills MHV amplitude \cite{Parke:1986gb} inspired a description in terms of 2D current algebra \cite{Nair:1988bq}. This idea was then generalised to $\cN = 4$ super Yang-Mills amplitudes\cite{Witten:2003nn,Berkovits:2004hg,Roiban:2004yf}, and was also found to calculate $\cN = 4$ conformal supergravity amplitudes\cite{Berkovits:2004jj}.  A similar worldsheet formula for $\cN = 8$ supergravity was found in\cite{Skinner:2013xp}. The spectra of these string models contains only field theory degrees of freedom. Cachazo, He and Yuan extended the formula of \cite{Skinner:2013xp} to a framework for calculating scattering of particles in arbitrary dimensions for a wide variety of theories in terms of a unified set of scattering equations\cite{Cachazo:2013gna,Cachazo:2013hca,Cachazo:2014xea}. We will refer to these equations as `general $d$ scattering equations'. In the CHY framework, tree-level amplitudes for different theories of massless particles are supported on the solutions of these scattering equations, which were first discovered in the context of ordinary string theory in \cite{Gross:1987kza, Fairlie:1972zz}. Worldsheet expressions of this form are related in a deep way to recursive formulae arising from BCFW\cite{Spradlin:2009qr,Farrow:2017eol,Nandan:2009cc,ArkaniHamed:2009dg}. 
 
The power of the spinor helicity formalism in four dimensions along with the generality of the scattering equation formalism were then combined to produce 4D ambitwistor string theory\cite{Geyer:2014fka,Bandos:2014lja}. This model gives a worldsheet description for the tree-level S-matrices of Yang-Mills theories and Einstein gravity which are valid for any number of supersymmetries and are supported on refined scattering equations which are graded by helicity degree. We will refer to these equations as `4D scattering equations'. In recent work we extended this formalism to include $\cN = 4$ conformal supergravity \cite{Farrow:2018yqf}. 

The scattering equations are understood in most detail at tree level, and this paper will consider only tree level amplitudes. There are extensions to the general $d$ scattering equations to calculate loop integrands\cite{Adamo:2013tsa, Geyer:2015bja,Cachazo:2015aol,Baadsgaard:2015twa}. The 4D scattering equations currently only support tree level computations, although some preliminary one loop expressions exist\cite{Farrow:2017eol}.

Formulae supported on scattering equations have been successful in representing and calculating abstract theoretical properties of amplitudes such as soft limits \cite{Lipstein:2015rxa, Schwab:2014xua, Cachazo:2015ksa, Adamo:2014yya, Geyer:2014lca}, collinear limits \cite{Nandan:2016ohb} and relations between the S-matrices of different theories \cite{Cachazo:2014xea, Nandan:2016pya}. Direct evaluation of amplitudes by solving the scattering equations is difficult, and some approaches solve the integrations by different methods\cite{Baadsgaard:2015voa, Baadsgaard:2015hia}. The equations have $(n-3)!$ solutions at $n$ points, and it is likely that finding all of these solutions analytically for generic kinematics is not possible. Calculating amplitudes in this framework then becomes a primarily numerical problem, which has been addressed only for the general $d$ equations. These equations can be reduced to a simplified polynomial form \cite{Dolan:2014ega}, which {\sc Mathematica}'s inbuilt {\tt NSolve} algorithm can solve numerically up to 9 points on a standard laptop. CHY provide an algorithm for finding individual solutions at higher points \cite{Cachazo:2013gna}, but there are difficulties finding all solutions in this way. Solutions which start out distinct at the start of the algorithm generically degenerate and the full set of solutions is not recovered at the end; details are discussed in this paper. The 4D equations do not currently have an equivalent simplified polynomial form and depend on a larger set of variables than the general $d$ equations, and as such {\tt NSolve} cannot solve them above 7 points. We overcome these difficulties with a Monte-Carlo solution finding algorithm which finds all solutions.

There exist published {\sc Mathematica} packages for evaluation of tree level amplitudes using BCFW\cite{Dixon:2010ik, Bourjaily:2010wh}, but to date there have been no equivalent packages for calculating amplitudes using the scattering equations either analytically or numerically.\cite{Sogaard:2015dba} provide an algorithm for calculation of amplitudes numerically without directly solving the scattering equations, but no explicit implementation is given. Available {\sc Mathematica} packages focus on Yang-Mills theory, and we provide the first explicit publicly available algorithms for calculating Einstein supergravity and $\cN = 4$ conformal supergravity amplitudes at tree level. 

The key purpose of this paper is to present the {\sc Mathematica} package {\tt treeamps4dJAF}, which explicitly calculates tree level amplitudes by solving the scattering equations. The analytical framework required for understanding the details of the 4D scattering equation formalism is covered, and we provide details of how to find full sets of solutions for a given set of numerical momenta and MHV degree by Monte Carlo algorithm. The package supports analytical computations in the MHV sector, and numerical computations in general MHV sectors for a wide variety of theories.

We hope that this paper, along with its associated package, will provide a computational tool that can serve for a number of different purposes. It can be seen as an lead-in to the scattering equation formalism those new to the field, as well as a general introduction to calculating amplitudes in {\sc Mathematica}. It can be used as a reference to check if new techniques for calculating amplitudes agree with previously known results for the theories currently supported, and a play box for discovering further theories that can be described in the 4D scattering equation formalism. We intend to update the package with integrands for new theories once they have been discovered.

The structure of the paper is as follows. We start with a review of the general~$d$ and 4D scattering equations and the integrands for currently supported theories in \sref{sec:review}. \Sref{sec:analytical} gives an overview of the analytical tools necessary to calculate amplitudes from the 4D scattering equation integral. Numerical methods for solving the equations outside of the MHV sector by Monte Carlo algorithm and for extracting component amplitudes from a superamplitude are detailed in \sref{sec:numeric}. We conclude and discuss further directions in \sref{sec:conc}. \Aref{sec:mathematica} introduces the {\sc Mathematica} package {\tt treeamps4dJAF} which gives a concrete implementation of all of the algorithms discussed in the paper and is included with the paper. The package's key functions are detailed in this appendix along with usage examples. Finally \aref{sec:analappen} gives detailed proofs of various properties of the 4D scattering equations, including those used in \sref{sec:analytical}.

\section{Review and Conventions}
\label{sec:review}
In this section we review relevant details of the general $d$ and 4D scattering equation formalism. For an $n$ point N$^{k-2}$MHV amplitude we define $\mathscr{N}$ to be the set of all of the particles, i.e. $\mathscr{N} := \{1, ... n\}$. We will refer to the following equations as the `general $d$ scattering equations' \cite{Cachazo:2013hca}

\begin{equation}
\begin{split}
 \sum_{j\in \mathscr{N}\atop j \neq i}\frac{k_i \cdot k_j}{s_i - s_j}=0,
\end{split}
\end{equation}

\noindent where $k_i$ are a set of $n$-point null momenta obeying momentum conservation, and $s_i$ are points on the Riemann sphere. In the general $d$ formalism, tree-level amplitudes can then be calculated as integrals of some integrand $f(s)$ over delta functions enforcing these equations,

\begin{equation}
\begin{split}
\cA_n = \int  \frac{d^n s}{\rm SL(2)}\prod_{i\in \mathscr{N}}{}'\delta\left( \sum_{j\in \mathscr{N}\atop j \neq i}\frac{k_i \cdot k_j}{s_i - s_j}\right) f(s).
\end{split}
\end{equation}

This type of integral arises from the calculation of a correlation function on a string worldsheet, and hence we will refer to the $s$ as worldsheet variables. The equations have the standard SL(2) symmetry arising from global conformal transformations on the worldsheet, which reduces the number of integration variables by three. To balance this, three delta functions are removed and a corresponding Jacobian factor added, as denoted by the notation $\prod_{i\in \mathscr{N}}{}'$. Any valid integrand $f$ must be covariant under this SL(2) symmetry, along with any other symmetries inherited from the corresponding amplitude. The number of integrations matches the number of delta functions, and hence this integral simply instructs us to sum the integrand times the relevant Jacobian factor over the $(n-3)!$ solutions to the scattering equations. In this way, calculating tree-level scattering for massless particles in many theories has been reduced to solving a set of algebraic equations. Further details can be found in \cite{Cachazo:2013gna,Cachazo:2013hca}.

In four dimensions we have additional structure due to the factorization of the Lorentz group SO(3,1) = SL(2)$_L\times$SL(2)$_R$,  and the corresponding MHV classification which splits amplitudes up into different sectors depending on the number of negative helicity particles scattered. These additional structures can be described in terms of the spinor-helicity formalism, reviewed in \cite{Elvang:2015rqa,Henn:2014yza}. To clarify the notation, we define angle and square brackets such that $\ket{i}\in$ SL(2)$_L$ and $\sket{i}\in$ SL(2)$_R$ describing the external momenta. Then $k_i = \sket{i}\bra{i} := \sket{i}\otimes \bra{i}$ is the explicitly null external momentum on leg $i$. We form antisymmetric SL(2) invariant brackets as $\ang{ij} := \det(\ket{i}\ket{j})$ and $\squ{ij} := \det(\sket{i}\sket{j})$.

We work with supersymmetry in an on-shell superspace via the Grassmann variables $\eta_i^A$, where $A$ is an R-symmetry index and $\eta_i^A$ are related to the supermomenta by $q_i^A = \ket{i}\eta_i^A$. Amplitudes in supersymmetric theories are most concisely described as a supersymmetry covariant superamplitude. This corresponds to scattering on-shell supermultiplets of the theory, and the superamplitude is a Grassmann expansion in terms of the $\eta^A_i$. The superamplitude will be a Grassmann function with some well-specified weight in $\eta$ variables $N_G$. Amplitudes where $N_G$ is not a multiple of $\cN$ are zero, and from this we define the Grassmann degree $k_G := \frac{N_G}{\cN}$ of the amplitude. Component amplitudes for individual states can be extracted by integrating against $N_G$ relevant $\eta$ variables. A review of these techniques can be found in \cite{Henn:2014yza, Elvang:2015rqa}, and the extension to non-maximal supersymmetry is discussed in detail in \cite{Elvang:2011fx}.

As an example consider $\cN = 4$ super Yang-Mills with supermultiplet $\Phi^{\pm} = g^{-}\eta_{1}\eta_{2}\eta_{3}\eta_{4}+\eta_{I}\eta_{J}\eta_{K}\psi^{-\,IJK}+\eta_{I}\eta_{J}\phi^{IJ}+\eta_{I}\psi^{+\, I}+g^{+}$. The MHV superamplitude $\cA(++--)$ has $k_G = 2$, and for maximal supersymmetry the positive and negative supermultiplet are the same and hence we can choose which legs are $+$ and $-$. For non-maximal SUSY the choice of assignment of superfields to each leg results in different superamplitudes. To extract a mixed fermion-gluon four point component amplitude, we read off the relevant $\eta$ from the supermultiplet, and integrate the superamplitude against these Grassmann variables;
 
\eq{
\cA(g^{-}\psi^{-\,IJK}g^+\psi^{+\, L}) = \int d \eta_1^1 d \eta_1^2 d \eta_1^3 d \eta_1^4 d \eta_2^I d \eta_2^J \eta_2^K d \eta_4^L \cA(++--).
}

The 4D scattering equations are a refinement of the general $d$ scattering equations which depend on the MHV degree as well as the number of particles\cite{Geyer:2014fka}, and calculate supersymmetry covariant expressions. They make use of spinor-helicity notation and are written in terms of the angle and square bracket spinors rather than momentum dot products. This introduces an extra variable $t_i$ into the system for each particle which is related to the little group scaling of the spinor variables. We will think of the $t_i$ as additional worldsheet variables, promoting each worldsheet coordinate from a point on the Riemann sphere to a point in $\mathbb{C}^{2}$. We will use different coordinates for the worldsheet variables as $\sigma_i =\frac{1}{t_i}\left(\begin{smallmatrix}1\\s_i\end{smallmatrix}\right) = \left(\begin{smallmatrix}\sigma_i^1\\ \sigma_i^2\end{smallmatrix}\right) \in \mathbb{C}^2$. Combining all of worldsheet variables together as a matrix $\sigma \in \mathbb{C}^{2\times n}$, we will work with minors of this matrix defined as $(ij) := \det(\sigma_i\sigma_j) = \frac{s_i-s_j}{t_i t_j} = \sigma_i^1\sigma_j^2 - \sigma_i^2\sigma_j^1$.


The external particles are now grouped into two different sets $L$ and $R$ with $L\sqcup R = \mathscr{N}$ to respect the MHV degree of the amplitude. For Yang-Mills theory and Einstein gravity, the left set will correspond to the set of negative helicity particles (or more generally, those particles that sit in the negative helicity super-multiplet of the theory), and the right set will correspond to the positive helicity particles (resp. positive supermultiplet). In these cases the MHV degree $k$ of the amplitude will be the size of the left set, $k := |L|$. More generally we can define an integrand in terms of some abstract left set, and indeed for example conformal supergravity amplitudes can have negative helicity particles in the left or the right sets, which we explain shortly.

We will refer to the following equations as the `4D scattering equations' \cite{Geyer:2014fka}. They are specified by an integer $n$ and a left set $L$.

\begin{equation}
\begin{split}
\label{eqn:scatteqns}
&\tilde{E}_l := \sket{l} - \sum_{r\in R}\frac{\sket{r}}{(lr)} = 0, \hspace{.5cm}l \in L\hspace{1.5cm}
E_r := \ket{r} - \sum_{l\in L}\frac{ \ket{l}}{(rl)} = 0\hspace{.5cm} r \in R \\
&\hspace{3cm}\diracd{2\times n}{SE^n_L} := \prod_{l\in L}\delta^2(\tilde{E}_l)
\prod_{r\in R}\diracd{2}{E_r}.
\end{split}
\end{equation}
 As in the general $d$ case, we can then write amplitudes for many different theories as integrals of some integrand $f$ over these scattering equation delta functions;

$$
\label{eqn:scattint}
\cA_{n,L} = \int \frac{d^{2\times n}\sigma}{GL(2)} \diracd{2\times n}{SE^n_L} f(\sigma, \ket{i}, \sket{i}).
$$

We will also analyse the integrands for some specific theories. Integrands are currently known for Einstein gravity and Yang-Mills theory with any valid number of supersymmetries\cite{Geyer:2014fka}, and for graviton multiplets in $\cN = 4$ conformal supergravity\cite{Farrow:2018yqf}. The supersymmetry structure is encoded by a set of fermionic delta functions of the form $\diracd{\cN}{\eta_l - \sum_{r\in R}\frac{\eta_r}{(lr)}}$, which can be thought of as scattering equations for the Grassmann on-shell superspace variables. In this paper we include the fermionic delta functions in the integrand, rather than as part of the scattering equations. The integrands for these theories are

\begin{eqnarray}
&\hspace{-.6cm}f_{\rm sYM}(\sigma,\ket{i},\sket{i},\eta_i) &:=\prod_{l\in L} \diracd{\cN}{\eta_l - \sum_{r\in R}\frac{\eta_r}{(lr)}} \frac{1}{\prod_{i \in N}(i\,i{+}1)}\nonumber\\
&\hspace{-.6cm}f_{\rm SUGRA}(\sigma,\ket{i},\sket{i},\eta_i) &:=  \prod_{l\in L}\diracd{\cN}{\eta_l - \sum_{r\in R}\frac{\eta_r}{(lr)}} \det{'}\cH \;\det{'}\tilde{\cH} \\
&\hspace{-.6cm}f_{\rm CSG}(\sigma,\ket{i},\sket{i},\eta_i) &:= \prod_{l\in L} \diracd{4}{\eta_l - \sum_{r\in R}\frac{\eta_r}{(lr)}}\prod_{l^- \in L\cap \Phi^-} \hspace{-.2cm}\cH_{l^-} \hspace{-.2cm}\prod_{l^+ \in L\cap \Phi^+} \hspace{-.2cm}\tilde{\cF}_{l^+}\hspace{-.2cm} \prod_{r^- \in R \cap \Phi^-}\hspace{-.2cm} \cF_{r^-}\hspace{-.2cm}\prod_{r^+ \in R\cap \Phi^+}\hspace{-.2cm}\tilde{\cH}_{r+} , \nonumber
\end{eqnarray}

\noindent  where $\det{}'$ is an instruction to remove one row and column from the matrices before taking the determinant. $\cH$ is the Hodges matrix and $\tilde{\cH}$ the dual Hodges matrix, defined as

\begin{align}
&\cH_{ll}:=-\sum_{l'\neq l\in L}\cH_{ll'},\hspace{.5cm} l \in L &\cH_{ll'}&:=\frac{\left\langle ll'\right\rangle }{\left(ll'\right)},\hspace{.5cm}l\neq l'\in L\nonumber\\
&\tilde{\cH}_{rr}:=-\sum_{r'\neq r}\tilde{\cH}_{rr'},\hspace{.5cm} r \in R &\tilde{\cH}_{rr'}&:=\frac{\left[rr'\right]}{\left(rr'\right)},\hspace{.5cm}r\neq r'\in R.
\end{align}

In conformal supergravity the left set $L$ does not correspond to the set of negative helicity superfields, which we denote $\Phi^-$ (with $\Phi^+$ the positive helicity superfields; $\Phi^-\sqcup\Phi^+=\mathscr{N}$). The left set $L$ of the scattering equations has $|L| = k_G$, where $k_G$ is the Grassmann degree of the superamplitude considered, and the choice of $L$ does not affect the amplitude. The factors $\cH$, $\cF$, $\tilde{\cH}$ and $\tilde{\cF}$ in conformal supergravity are generalisations of the gravitational inverse soft factor outside of the MHV sector. The $\cH$ factors are the diagonal elements of the Hodges matrices, and the $\cF$ are given by

\begin{align}
\tilde{\cF}_{l}:=\sum_{r<r'\in R }\frac{\left[rr'\right]\left(rr'\right)}{\left(lr\right)^{2}\left(lr'\right)^{2}} & &
\cF_{r}:=\sum_{l<l'\in L}\frac{\left\langle ll'\right\rangle \left(ll'\right)}{\left(rl\right)^{2}\left(rl'\right)^{2}}.
\end{align}

As shown in \cite{Lipstein:2015rxa}, there are $A(n,k) = \ang{\begin{smallmatrix}n-3\\k-2\end{smallmatrix}}$ solutions to the $n$-point N$^{k-2}$MHV scattering equations. $\ang{\begin{smallmatrix}n\\k\end{smallmatrix}}$ are the Eulerian numbers\cite{Euleriannumbers}, tabulated in \fref{fig:solcount}. The inductive proof of the number of solutions finds that $A(n,k)$ can be broken down by taking either a left set particle soft to reduce to an amplitude of the form $A(n-1,k-1)$ or a right set particle soft to reduce to $A(n-1,k)$. Each of these solutions then has a given multiplicity such that 

\begin{equation}
\label{eqn:solcount}
A(n,k) = (n-k-1)A(n-1,k-1) + (k-1)A(n-1,k),
\end{equation}

\noindent with one solution in the MHV and \MHVB sectors for each $n$, so that $A(n,2) = A(n,n-2) = 1$.

\section{Calculating Amplitudes with the 4D Scattering Equations}
\label{sec:analytical}

In this section we describe the analytical results necessary for solving the scattering equations to calculate amplitudes given a certain integrand. These techniques will then be used explicitly to find amplitudes numerically and analytically in the package {\tt treeamps4dJAF}. Detailed proofs are given in \aref{sec:analappen}. 

As shown in \eref{eqn:scatteqns}, at $n$ points there are $2n$ 4D scattering equations depending on $2n$ worldsheet $\sigma$ variables. There is a GL(2) symmetry on the worldsheet which acts as inhomogenously on the left and right set of worldsheet coordinates for $G \in$ GL(2) as

\eq{
\label{eqn:GL2}
 \sigma_l \rightarrow G \,\sigma_l,\hspace{1cm} l \in L \hspace{2cm} \sigma_r \rightarrow \frac{G}{\det G}\, \sigma_r, \hspace{1cm}r \in R. 
}
 
 \noindent Under this GL(2) any minor of the form $(lr)$ remains invariant, and hence the scattering equations are invariant. Fixing this gauge symmetry leaves $2n-4$ remaining degrees of freedom. Generally in this work we will restrict to gauge transformations specified by two particle labels $i, j$ which fix $\sigma_i = \left(\begin{smallmatrix}1 \\0 \end{smallmatrix}\right)$ and $\sigma_j = \left(\begin{smallmatrix}0 \\1 \end{smallmatrix}\right)$, and refer to this operation as `gauge-fixing particles $i$ and $j$'. As the GL(2) transformations act inhomogeneously on the left and right set, we can only fix either two left or two right particles this way. The details of gauge-fixing and the symmetries of the equations are explained in \aref{sec:syms}.

The system now appears to be over specified, and we must remove four equations. Two spinor equations $i$ and $j$ either from the left set or from the right set can be reduced to a momentum conserving delta function on support of the other scattering equations, as proved in \aref{sec:gfix}. We then refer to having `deleted particles $i$ and $j$', and the remaining equations are a well-specified set of $2n-4$ equations in $2n-4$ variables. 

The number of integrations in \eref{eqn:scattint} is the same as the number of delta functions, and hence the integrations instruct us to sum over all of the solutions of the scattering equations. Deleting equations $l, l' \in L$ we can calculate the Jacobian of the remaining equations to solve the delta function integrals as follows

\eq{
\cA_{n,L} = \int \frac{d^{2\times n}\sigma}{GL(2)} \diracd{2\times n}{SE^n_L} f(\sigma) = \delta^4(P) \sum_{\sigma_{\rm sol} \in {\rm solutions}} \frac{ f(\sigma_{\rm sol})}{\ang{ll'}^{-2}\det(J^{n\,ll'}_{L}(\sigma_{\rm sol}))} .
}

\noindent where $J^{n}_L$ is the Jacobian of the scattering equations with respect to the sigma variables, and the superscript $l,l'$ refers to removing four rows and columns corresponding to the two particle labels from the matrix. Details of the Jacobian to the scattering equations are explained in \aref{sec:jac}. It is now a well-formulated problem to solve the scattering equations and sum a theory dependent integrand $f$ over the full set of solutions to produce an amplitude.

Calculating MHV amplitudes is generally more simple than calculating N$^{k-2}$MHV amplitudes, and we find this simplicity to be reflected in the structure of the 4D scattering equations. Analytical solutions to the 4D scattering equations are not known for general MHV degree, but in the MHV sector we can construct analytical solutions. First consider the case where the left set is $L = \{1,2\}$. Then the MHV equations become

\begin{equation}
\begin{split}
\diracd{2\times n}{SE^n_{\{1,2\}}} = \diracd{2}{|1] - \sum_{r\in R}\frac{|r]}{(1r)}}
\diracd{2}{|2] - \sum_{r\in R}\frac{|r]}{(2r)}}
\prod_{r \in R}\diracd{2}{\ket{r} - \frac{ \ket{1}}{(1r)}+\frac{ \ket{2}}{(2r)}}
\end{split}
\end{equation}

The most obvious choice of equations to remove in this case is the two left set equations which become the overall momentum conservation delta function as detailed in \aref{sec:gfix}, and we recover a Jacobian of $\ang{12}^2$. Similarly we gauge-fix particles 1 and 2 to the identity in the Grassmannian, and arrive at the following form for each delta function of the right-set equations, which are solved by a Schouten identity;
\begin{equation}
\begin{split}
&\diracd{2}{\ket{r} - \frac{ \ket{1}}{\sigma_r^2}-\frac{ \ket{2}}{\sigma_r^1}} 
=  \frac{\ang{12}^3}{\ang{1r}^2\ang{2r}^2}\diracd{}{\sigma_r^1 - \frac{\ang{12}}{\ang{r1}}}\diracd{}{\sigma_r^2 - \frac{\ang{12}}{\ang{r2}}}, \hspace{.3cm}r \in R.
\end{split}
\end{equation}

The full MHV solution along with its minors and the associated expression for the Jacobian of the delta functions is

\begin{equation}
\begin{split}
&\hspace{3.7cm}\sigma_{\text{MHV}} = \left(\begin{matrix}
1 & 0 & \frac{\ang{12}}{\ang{31}} &...& \frac{\ang{12}}{\ang{n1}}\\
0 & 1 & \frac{\ang{12}}{\ang{32}} &...& \frac{\ang{12}}{\ang{n2}}
\end{matrix}\right)\\
&\hspace{2.2cm}\diracd{2\times n}{SE^n_{\{1,2\}}} = \ang{12}^2 \prod_{r \in R} \frac{\ang{12}^3}{\ang{1r}^2\ang{2r}^2}  \diracd{4}{P} \diracd{2n-4}{\sigma - \sigma_{\text{MHV}}}\\
&(12)_{\text{MHV}} = 1  \hspace{.5cm}
(rr')_{\text{MHV}} = \frac{\ang{12}^3\ang{rr'}}{\ang{1r}\ang{1r'}\ang{2r}\ang{2r'}} \hspace{.5cm}
(1r)_{\text{MHV}} =\frac{\ang{12}}{\ang{r2}}   \hspace{.5cm}
(2r)_{\text{MHV}} = \frac{\ang{12}}{\ang{1r}}.
\end{split}
\end{equation}
The Jacobian as calculated this way is in agreement with the calculation from \aref{sec:jac}, where we also consider the Jacobian for higher MHV degree.

Finding analytical solutions for generic kinematics outside of the MHV sector is currently an unsolved problem, apart from at 6 points NMHV where the scattering equations have four solutions. These solutions are found for the general $d$ equations for $d = 4$ in \cite{Weinzierl:2014vwa}. Abel's theorem states that there is no algebraic solution in terms of $n$th roots to a general polynomial equation of degree five or higher with arbitrary coefficients \cite{Abeltheorem}. To find analytical N$^{k-2}$MHV solutions above 6 points NMHV some underlying structure would have to exist within the coefficients of the equations, otherwise general analytical solutions are excluded by Abel's theorem. Full sets of solutions can be found analytically for some specific choices of momenta\cite{Kalousios:2013eca}.

Due to these difficulties in finding analytical solutions, calculating amplitudes by solving the scattering equations outside of the MHV sector is primarily a numerical problem, which we address in \sref{sec:numeric}.

One key strength of the scattering equation formalism is that once a full set of solutions are known, amplitudes can be calculated in any theory at relatively small computational cost. The relevant integrand is chosen, and the only necessary operation is to sum over the solutions. Solutions to the scattering equations are graded only by MHV degree and not by a specific choice of left set. This implies that there must exist some transformation on the worldsheet which can map an integrand supported on scattering equations for one left set into an integrand for a different left set of the same length. For example such a mapping will allow calculation of all 6 point NMHV gluon amplitudes in Yang Mills theory with only one solution to the scattering equations; eg. $\cA(+-+-+-)$ with left set $L = \{2,4,6\}$ and $\cA(---+++)$ with left set $L = \{1,2,3\}$.

The following is an explicit co-ordinate transformation on the worldsheet which swaps two particles between the left and right sets of the scattering equations.\footnote{We thank Paul Heslop for suggesting this transformation} We single out two legs $l_0 \in L$ and $r_0 \in R$ which which will be swapped. 

\begin{align}
\label{eqn:permtrans}
	\sigma_{l_0}\rightarrow \sigma'_{l_0}&= \sigma_{l_0}\frac1{(l_0r_0)},&\quad \sigma_{r_0}\rightarrow \sigma'_{r_0} &= \sigma_{r_0}\frac1{(r_0l_0)},\nonumber\\
\sigma_{l}\rightarrow \sigma'_{l}&= {\sigma_{l}}\frac{(ll_0)}{(lr_0)}\hspace{.5cm}l\neq l_0\in L,\hspace{.5cm}&\quad\sigma_{r}\rightarrow \sigma'_{r}&= {\sigma_{r}}\frac{(rr_0)}{(rl_0)} \hspace{.5cm} r\neq r_0\in R.
\end{align} 
Under this transformation we find that the scattering equation integrand transforms as

\begin{equation}
\begin{split}
\label{eqn:scattinttrans}
&\int \frac{d^{2\times n}\sigma}{GL(2)} \diracd{2\times n}{SE^n_L} f(\sigma) \rightarrow  \\
&\hspace{1cm}\int \frac{d^{2\times n}\sigma}{GL(2)} \diracd{2\times n}{SE^n_{L'}} f'(\sigma) = 
\int \frac{d^{2\times n}\sigma}{GL(2)} \diracd{2\times n}{SE^n_{L'}}\prod_{l \neq l_0\in L}\frac{(ll_0)^2}{(lr_0)^2}\prod_{r \neq r_0\in R}\frac{(rr_0)^2}{(rl_0)^2} \frac{f(\sigma')}{ (l_0r_0)^{8}}
\end{split}
\end{equation}
 where $L'$ has $l_0$ swapped with $r_0$. We now have an explicit transformation for how to calculate a new integrand $f'$ for a swap of the choice of left sets for the scattering equations. Details of the transformation under this mapping are given in \aref{sec:perms}. Repeatedly applying this transformation can be used to reassign any left set.

\section{Numerical Methods}
\label{sec:numeric}

The 4D scattering equations can be thought of as $2n-4$ equations with variables in the Grassmannian Gr(2,~$n$), as explained in \aref{sec:syms}. The equations are parametrised by a set of spinors obeying momentum conservation. A solution to the scattering equations at $n$ points N$^{k-2}$MHV will be a mapping from external data to $\ang{\begin{smallmatrix}n-3\\k-2\end{smallmatrix}}$ points in Gr(2,~$n$). In the MHV case we give this mapping analytically in \sref{sec:analytical}, but finding analytical solutions for $k > 2$ is complicated due to the combinatorially increasing number of solutions. A well-specified problem is to provide explicit numerical momenta, which will usually be randomly sampled, and to then solve the resulting equations numerically. CHY provide an inverse-soft type algorithm for finding individual numerical solutions to the general $d$ equations \cite{Cachazo:2013gna}, but there are difficulties in constructing the full set of solutions in this way which we discuss in \sref{sec:inversesoft}. We provide an explicit algorithm which given a set of numerical momenta samples random numerical points in Gr(2,~$n$) to find solutions to the equations stochastically. Algorithms of this type are known as Monte Carlo algorithms. Monte Carlo methods in high energy physics are well studied \cite{Lepage:1977sw,Alwall:2014hca} and their application to solving non-linear algebraic equations is straightforward. 

With enough computing power and time, any non-linear system of equations can be solved by Monte Carlo algorithm. The two key questions to address are when to stop the algorithm, and what distribution to sample the initial guess points from. The scattering equations are well-suited for solution in this way because the number of solutions is known, which gives a clear stopping condition. We address the sampling question in \sref{sec:montecarlo}. Finding a set of solutions this way is stochastic and can take a long time, with time complexity now distributed as a random variable which depends on $n$ and $k$. The expectation of the time complexity increases as $n$ increases and as $k$ moves towards $\lfloor \frac{n}{2}\rfloor$. One advantage of the 4D formalism that makes it better suited for solution by Monte Carlo algorithm is that the $(n-3)!$ solutions are broken down into Eulerian numbers of solutions, tabulated in \fref{fig:solcount}. This means that the algorithm can stop after finding a smaller number of solutions than in general dimensions.

Once a full set of numerical solutions along with the corresponding momenta and left set are known for a given number of points and MHV degree, they can then be used to calculate amplitudes in different theories for a selection of different external states by substituting different integrands into the sum over solutions. Solutions up to 12 points NMHV and 9 points in all MHV sectors are currently accessible to the algorithms of {\tt treeamps4dJAF}, and we tabulate full solution sets with rational external data for these cases in the accompanying data file {\tt SolutionLookupTable.csv}.

Tree-level amplitudes are all rational functions of external momenta, and hence for rational numerical external data they will be a rational number. The solutions to the scattering equations are in general not rational numbers, but given a set of rational kinematics we can calculate to very high precision at relatively low computational cost via deterministic algorithm once all solutions are known. It is then possible rationalize to the closest rational number to give exact numerical results for the amplitude. We provide support for this type of calculation in {\tt treeamps4dJAF}.

\begin{figure}[h]
\centering
\begin{tabular}{cc}
$n$ & $\ang{\begin{smallmatrix}n-3\\k-2\end{smallmatrix}}$ \\
4 & 1\\    
5 & 1 1	\\    
6 & 1 4 1\\ 
7 & 1 11 11 1\\   
8 & 1 26 66 26 1\\
9 & 1 57 303 302 57 1 \\
10 & 1 120 1191 2416 1191 120 1 \\
\end{tabular}
\caption{Eulerian numbers of solutions to the 4D scattering equations}
\label{fig:solcount}
\end{figure}

\subsection{Difficulties with CHY's Inverse Soft Algorithm}
\label{sec:inversesoft}
One proposed algorithm to find numerical solutions to the general $d$ equations is that of CHY, which takes one of the momenta soft with parameter $\epsilon$ to reduce the equations from $n$ points down to $n-1$ points\cite{Cachazo:2013gna}. The soft parameter is then reintroduced, and the soft equation at $O(\epsilon)$ is solved for each of the $(n-4)!$ solutions to the $n-1$ point equations. The solutions then have a multiplicity of $n-3$, and these points are input back into the system with $\epsilon$ moving slowly up from 0~to~1. As this algorithm involves slowly bringing the soft parameter back to the full $n$ point system, it is referred to as an inverse soft algorithm. $(n-3)!$ solutions to the $n$ point equations will be found in this way, but there is no guarantee that all of these solutions will be distinct, and hence they will not necessarily cover the full solution space.

The inverse soft algorithm is based on an inductive argument for counting the number total number of solutions, and we can extend it to the 4D case using the analogous 4D solution counting argument, which is reviewed in \sref{sec:review}. We take one soft parameter $\epsilon$ for a left set particle $1 \in L$ so that $\ket{1} \rightarrow \epsilon \ket{1}$, and a further parameter $\tilde{\epsilon}$ for a right set particle $n\in R$ so that $\sket{n} \rightarrow \tilde{\epsilon} \sket{n}$. The 4D scattering equations become 

\eqs{
& \hspace{-.4cm}\sket{1} - \sum_{r\neq n \in R}\frac{\sket{r}}{(1r)} - \tilde{\epsilon} \frac{\sket{n}}{(1n)}= 0\hspace{3.2cm}\ket{n} - \epsilon\frac{ \ket{1}}{(n1)} - \sum_{l\neq 1\in L}\frac{ \ket{l}}{(nl)} = 0\\
&\hspace{-.4cm}\sket{l} - \sum_{r\neq n \in R}\frac{\sket{r}}{(lr)} - \tilde{\epsilon} \frac{\sket{n}}{(ln)}= 0, \hspace{.5cm}l\neq 1 \in L\hspace{1.cm}
\ket{r} - \epsilon\frac{ \ket{1}}{(r1)} - \sum_{l\neq 1\in L}\frac{ \ket{l}}{(rl)} = 0\hspace{.5cm} r\neq n \in R.
}

\noindent We can see that worldsheet variable for the particle in the soft limit decouples, and the equations reduce to $n-1$ point N$^{k-2}$MHV equations when $\epsilon = 1, \tilde{\epsilon} \rightarrow 0$ and $n-1$ point N$^{k-3}$MHV equations when $\tilde{\epsilon} = 1, \epsilon \rightarrow 0$. Evaluated on the solution to the lower point equations, the remaining equation for the particle that decoupled gives the multiplicity for each solution, as shown in \eref{eqn:solcount}.
 
\begin{figure}[h]
\centering
       \includegraphics{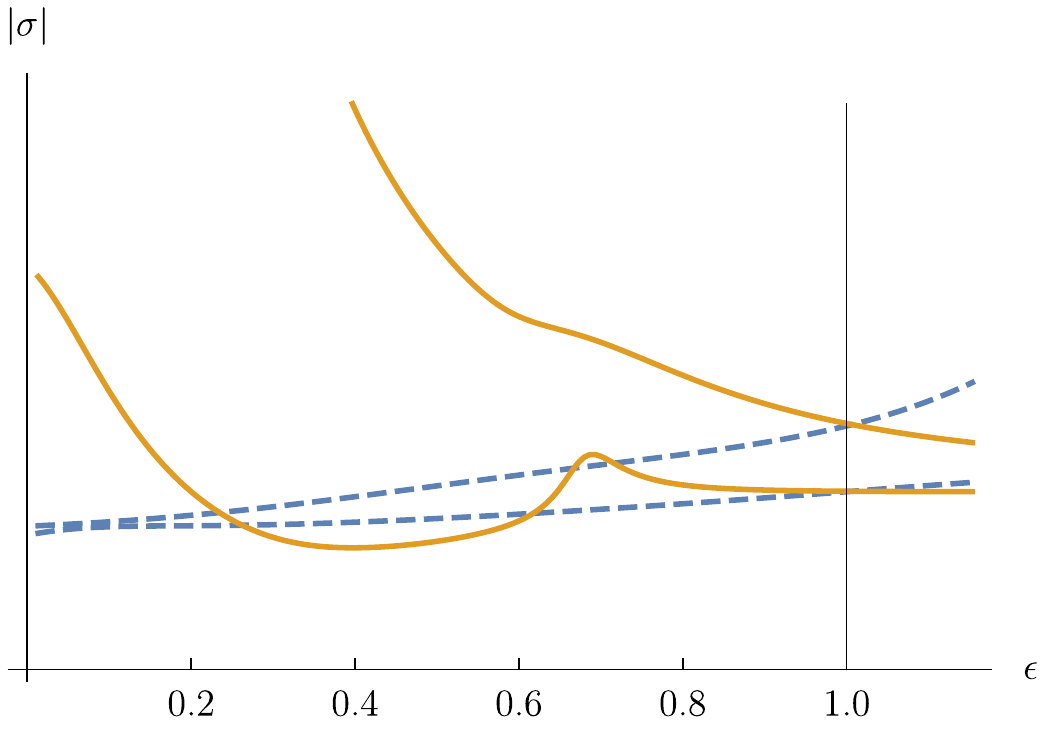}
    \caption{Convergent solutions at 6 points NMHV using CHY's inverse soft algorithm. Orange solid lines are solutions coming from 5 points \MHVB and blue dashed lines are from 5 points MHV.} 
    \label{fig:convergentsols}
\end{figure}

As detailed in \cite{Cachazo:2013gna} this method is sufficient to produce individual solutions for a specific $n$ and $k$, but we find difficulties when trying to construct all of the solutions in this way. In four dimensions we find that two of the different solutions constructed from a lower point amplitude can converge to the same higher point solution, as shown at 6 points NMHV in \fref{fig:convergentsols}. Hence the maximum number of solutions this algorithm can find is $\ang{\begin{smallmatrix}n-3\\k-2\end{smallmatrix}}$, and generically it does not find all of the solutions.

MHV and \MHVB solutions are known analytically, and the first non MHV case is at 6 points NHMV, with four solutions. The $\epsilon$ soft limit gives a 5 point \MHVB amplitude, and the $\tilde{\epsilon}$ soft limit produces a 5 point MHV amplitude. The soft limit equations both have two solutions. \fref{fig:convergentsols} describes the norm of the MHV solutions as they evolve from $\epsilon= 0$ up to $\epsilon= 1$ in blue, and the \MHVB solutions from $\tilde{\epsilon}= 0$ up to $\tilde{\epsilon}= 1$ in yellow.

We define a matrix norm on the solutions by taking the standard norm on $\mathbb{C}^n$, after flattening the $2\times4$ worldsheet matrix with gauge-fixed rows deleted down to $\mathbb{C}^8$. Lines crossing on the plot in this description is not sufficient for the solutions to be equal, and indeed where the lines cross for $\epsilon \in [0,1]$ the solutions are not equal. At $\epsilon = 1$ the solutions converge. Interestingly when we continue the algorithm to run for $\epsilon$ slightly larger than the 1 we see that the solutions separate again.

In the case shown in \fref{fig:convergentsols}, both of the solutions from 5 points MHV collide separately with two individual solutions from 5 points \MHVB. In general for different randomly selected numerical momenta at 6 points NMHV we find zero, one or two pairs of solutions converging. The number of pairs that converge appears to be random based on the selection of random momenta. Solution crashing is a ubiquitous phenomenon; it occurs for nearly all choices and the difficulty is rather to find cases where the solutions do not crash than to find cases where they do.

These difficulties with the inverse soft algorithm inspire developing new methods, and we solve this problem using a Monte Carlo algorithm.

\subsection{Monte Carlo Algorithm}
\label{sec:montecarlo}

The Monte Carlo equation solving algorithm in {\tt treeamps4dJAF} is implemented via \linebreak{\tt NSolveMonteCarlo}. Many initial random points are sampled from a distribution described below, and chosen as the initial conditions for a {\tt FindRoot} calculation. These initial calls to {\tt FindRoot} run for many iterations, and stop after only one digit of precision is met. Most initial guess points will be far from a solution, and will not converge to 1 digit of precision by the specified number of iterations. Those which do not converge are discarded, and the ones that do converge go back into {\tt FindRoot} up to a higher precision. These points are now solutions, which are compared with a list of all currently found solutions and duplicates are discarded. The algorithm stops when a suitable stopping condition is met, which may be after a specified amount of time or number of iterations, or when enough solutions are found. A pseudo-code for this algorithm is

\begin{algorithmic}
\Function{\tt NSolveMonteCarlo}{\var{equations}, \var{variables}}
\State Compile equations and Jacobian down to C code for faster evaluation
\vspace{.2cm}\While{a stopping condition is not met}	
\State Sample 100 initial solution points from a specified distribution 
\vspace{.2cm}\State Run {\tt FindRoot} on each point. Stop after 1000 iterations, or when a point \State solves the equations to 1 digit precision
\vspace{.2cm}\State Run {\tt FindRoot} to higher precision for points that solved the equations
\vspace{.2cm}\State Compare solutions to a specified precision and discard any duplicate 
\State solutions found
\EndWhile

\vspace{.2cm}\State Return the solutions
\EndFunction
\end{algorithmic}

\begin{figure}[h]
\centering
    \includegraphics{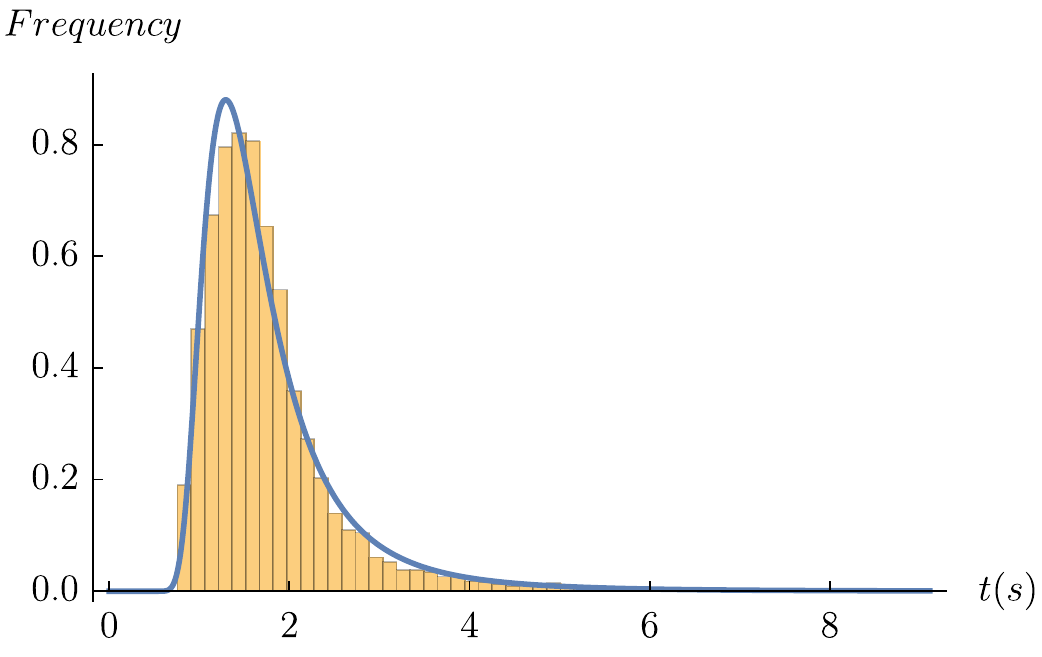}
    \caption{Histogram of timings for finding solutions for 7559 sets of solutions to the 4D scattering equations at 6 pts NMHV using {\tt NSolveScatteringEquations4D}. The blue curve is {\tt FrechetDistribution[3.7,1.6,-0.2]}, as a best fit by {\sc Mathematica}.} 
    \label{fig:timingshistogram}
\end{figure}

 {\tt NSolveMonteCarlo} is tailored to the 4D scattering equations in the function \linebreak{\tt NSolveScatteringEquations4D}. This specifies the stopping condition to be when all of the Eulerian numbers of solutions are found, and selects an appropriate distribution to sample the random points from. \fref{fig:montecarlodata} gives a statistical analysis of the time complexity of the algorithm\footnote{All timings were calculated on a Linux desktop computer with 3.30GHz Intel(R) Core(TM) i7-5820K processor, and vary depending on what other processes were running during evaluation of the algorithm.} and of the distribution of solution points for all currently accessible $n$ and $k$. \fref{fig:timingshistogram} expands on this analysis for 6 points NMHV, giving a histogram of the timings. Timings are positively skewed with a similar shape for other $n$ and $k$. Based on this, the algorithm can currently handle at most around 500 total solutions. Accessing the next cases would require around 1000 solutions, at 10 points N$^2$MHV and 13 points NMHV.

\begin{figure}[h]
\centering
\begin{tabular}{|c|c|c|c|c|c|c|c|c|}
\hline
$n$ & $k$ & $\ang{\begin{smallmatrix}n-3\\k-2\end{smallmatrix}}$& $N_{\rm runs}$ & $N_{\sigma}$ &$\gamma_\sigma$ & MAD$_\sigma$ & $\overline{t}$ & STD$_t$  \\
\hline
6 & 3 & 4 & 7559 & 483776 & 0.448 & 0.462 &  1.82 s &  1.23 s \\
7 & 3 & 11 & 1047 & 230340 & 0.433 & 0.448 &  12.5 s &  12.2 s \\
8 & 3 & 26 & 1001 & 624624 & 0.443 & 0.461 &    1. min &  50.1 s \\
9 & 3 & 57 & 57 & 90972 & 0.408 & 0.427 &  22.9 min &  15.8 min \\
10 & 3 & 120 & 9 & 34560 & 0.423 & 0.427 &  4.33 hr &  3.65 hr \\
11 & 3 & 247 & 2 & 17784 & 0.508 & 0.508 &  10.7 hr &  1.57 hr \\
7 & 4 & 11 & 1001 & 220220 & 0.482 & 0.485 &  39.8 s &  1.49 min \\
8 & 4 & 66 & 53 & 83952 & 0.439 & 0.447 &   31. min &  29.9 min \\
9 & 4 & 302 & 2 & 16912 & 0.397 & 0.403 &  28.7 hr &  2.67 hr \\
8 & 5 & 26 & 328 & 204672 & 0.539 & 0.533 &  3.45 min &  3.02 min \\
9 & 6 & 57 & 51 & 81396 & 0.665 & 0.642 &  21.2 min &  17.6 min \\
10 & 7 & 120 & 17 & 65248 & 0.698 & 0.702 &  59.8 min &  34.9 min \\
11 & 8 & 247 & 2 & 17784 & 0.627 & 0.61 &  11.5 hr &  7.81 hr \\
12 & 9 & 502 & 1 & 20080 & 1.71 & 1.76 &  22.3 hr &  -  \\
\hline
\end{tabular}

    \caption{Statistical summary of distribution of solutions to 4D scattering equations and timings of {\tt NSolveMonteCarlo} algorithm. $N_{\rm runs}$ is the number of different set of numerical momenta used, $N_\sigma = (4n-8) \ang{n-3 \atop k-2}N_{\rm runs}$ is the number of solution points, $\gamma_\sigma$ is the scaling parameter of the fitted Cauchy distribution and MAD$_\sigma$ is the median absolute deviation.  $\overline{t}$ is the average time and STD$_t$ is the standard deviation. Note that when $N_{\rm runs}$ is small solution point statistics may not be reliable even though $N_\sigma$ is large, as they come from a small number of different choices of numerical momenta.} 
    \label{fig:montecarlodata}
\end{figure} 

\begin{figure}[h]
\centering
    \includegraphics{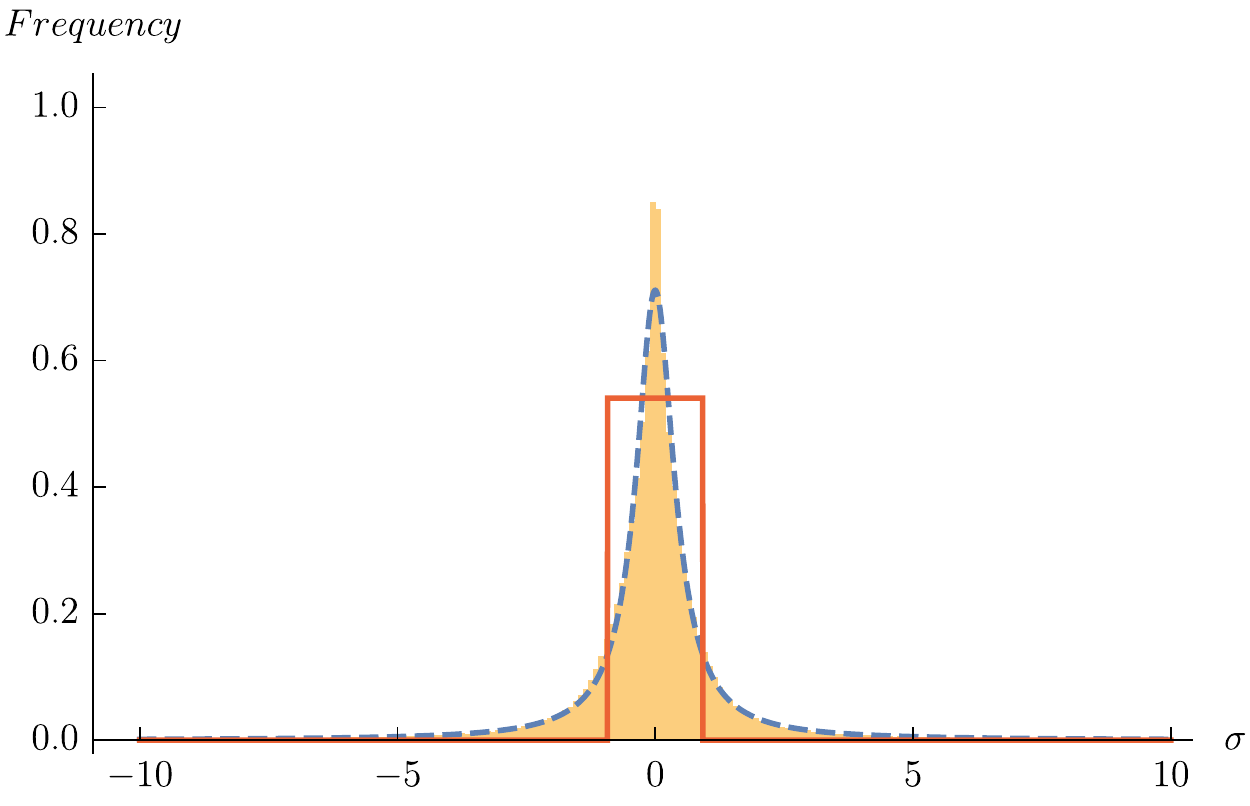}
    \caption{Histogram of 43776 solution points at 6 points NMHV taken from 7559 different random momenta. The blue dashed curve is {\tt CauchyDistribuion[0,0.447]}, and the red solid curve is an approximation with the uniform distribution $\mathcal{U}([-0.894,0.894])$.} 
    \label{fig:solpointshistogram}
\end{figure} 

We now analyse the sampling of the initial points, and explain which distribution to sample from for most efficient results. We solve momentum conservation in terms of the spinors $\sket{1}$ and $\sket{2}$, and delete equations 1 and 2 so that these spinors do not appear in the equations to be solved. This way every random numerical spinor generated is unconstrained, and hence comes from the same uniform distribution. Note that the statistics of the solution points is gauge dependent, and as we always gauge fix particles 1 and 2 it is not possible to make a direct comparison between data points with $k = k'$ and with $k = n-k'$, even though these cases at first glance should be symmetrical.

We perform a statistical analysis at 6 points NMHV, as guided by our analytical understanding in the MHV sector. A solution at 6 points NMHV can be considered as a matrix in $\mathbb{C}^{2\times 4}$ after removing gauge-fixed columns. We project this down to a vector of real numbers in $\mathbb{R}^{16}$, and hence collect 48 real numbers for each set of numerical momenta which we refer to as `solution points'. We then run the algorithm for statistically many sets of random numerical momenta, and collect all of the solution points together into one dataset. A histogram of this data is plotted in \fref{fig:solpointshistogram}. The data are best fitted by a Cauchy distribution, for which the probability density function has the form

\eq{
\mathbb{P}(x; x_0, \gamma) = \frac{1}{\pi \gamma}\left(\frac{\gamma^2}{(x-x_0)^2 + \gamma^2}\right).
}

The data are symmetrically distributed around 0, and hence we see that the location parameter $x_0 = 0$. This leaves only one remaining parameter $\gamma$, which is tabulated for some cases in \fref{fig:montecarlodata}. Apriori we would expect $\gamma$ be a function of $n, k$ and the size of the uniform distribution from which the random momenta are sampled. We can use some intuition from the analytical result in the MHV sector to reduce this down to just $\gamma = \gamma(n, k)$. Solution points in the MHV sector have a form such as ${\rm Re}(\frac{\ang{12}}{\ang{1r}})$, as shown in \sref{sec:analytical}. Hence the size of the uniform distribution from which the random events are sampled will divide out statistically, and will not affect the distribution of the solution points. This independence is guaranteed for higher MHV degrees as solutions has mass dimension 0, and spinors have mass dimension $\frac{1}{2}$. Performing a statistical analysis of numerical solutions to the scattering equations outside of the MHV sector also verifies these properties, for example as shown in \fref{fig:montecarlodata} where different values for $\gamma$ are tabulated. Our data in \fref{fig:montecarlodata} show that $\gamma$ is insensitive to $k$ also. We find this independence of $\gamma$ on $n$ and $k$ to be intriguing and counter-intuitive.

From this analysis, it seems clear that we should sample the initial points from the relevant Cauchy distribution. However, very many initial points must be sampled for each iteration of the algorithm, and sampling points from Cauchy distribution is significantly slower than sampling from a uniform distribution. We find in practice that it is more efficient to approximate this Cauchy distribution with a uniform distribution. To approximate a normal distribution with a uniform distribution, one could choose a symmetric range with a width which is a small multiple of the standard deviation, eg. $\mathcal{U}([-2\sigma,2\sigma])$. This poses a problem with the Cauchy distribution as it has infinite standard deviation. We use instead the median absolute deviation (MAD), which is roughly equivalent to the $\gamma$ parameter of the Cauchy distribution, and is tabulated for different $n$ and $k$ in \fref{fig:montecarlodata}. To solve the equations for a given $n$ and $k$, {\tt NSolveScatteringEquations4D} uses the tabulated MAD values from \fref{fig:montecarlodata} to sample initial points from  $\mathcal{U}([-2 \,{\rm MAD},2 \,{\rm MAD}])$. 

When finding solutions by Monte Carlo algorithm, a large percentage ($\sim\frac{3}{4}$) of the solutions tend to be found comparatively quickly, and it can take a long time to find the remaining solutions. It is possible that sampling a certain percentage of the solutions from the approximated uniform distribution and then sampling the remaining solutions from the relevant Cauchy distribution could help to overcome this problem. We leave this for future work.

\subsection{Efficient Component Amplitude Extraction}
\label{sec:grassmann}

Extracting individual component amplitudes from a super amplitude written as an expansion in Grassmann parameters is a well specified and well understood operation, as explained in \sref{sec:review}. Explicitly evaluating these calculations on the computer is not as simple as the operation itself might suggest. The most na\"{i}ve application is to expand out the Grassmann super amplitude in terms of each of its factors, and to then throw away any of the terms which are equal to zero either due to $\eta^2 = 0$ or because do not match the integration measure. This algorithm is prohibitive in terms of memory usage in the computer. For example a product of $m$ factors each with a sum of $m$ terms of Grassmann numbers will result in a total of $m^m$ terms when expanding out naively, nearly all of which are zero. This motivates finding a more efficient algorithm.

In the scattering equation formalism, the Grassmann delta functions are always written in the form $\prod_{l\in L} \diracd{\cN}{\eta_l - \sum_{r\in R}\frac{\eta_r}{(lr)}}$. We can see that integrals of this expression over a subset of the $\eta$ variables of dimension $k\cN$ is a special case of a Grassmann integral of the form $I_G := \int d^m \eta\prod_{i = 1}^m (\sum_{j = 1}^{m}A_{ij}\eta_j)$; in our real use case the matrix $A$ has some components given by worldsheet minors, and the rest zero. Grassmann integrals of this type can be evaluated in an especially neat analytical form, and we find that $I_G = \det(A)$. Numerical computation of determinants is generally implemented with an algorithm of time complexity around $O(m^3)$, and hence we can extract Grassmann components in a very efficient way using this formula. Grassmann integration of functions of this form then consists simply of assigning the correct values to the matrix $A$ and calculating its determinant. Any product of Grassmann delta functions takes this form, and specifically the fermionic delta functions of the scattering equations, as well as many other standard representations for superamplitudes.

We can improve the RAM and time efficiency of this algorithm further for numerical computations by storing the matrix as a sparse array, where all elements are assumed to be zero unless they are specified (in our usage case as worldsheet minors). We can then hold the evaluation of the determinant until explicit numerical values are substituted in for the worldsheet minors, and then calculate the determinant of a numerical matrix. Calculation in this way is very efficient, saving the need to store large intermediate analytical expressions in the RAM.

\section{Conclusions}
\label{sec:conc}

In this paper we provide the necessary tools for solving the 4D scattering equations analytically in the MHV sector and numerically for higher MHV degree. We outline a Monte Carlo algorithm for solving the scattering equations numerically, and an algorithm for extracting component amplitudes from Grassmann delta functions efficiently. This allows computation of tree-level amplitudes in Yang-Mills theory and Einstein gravity with any number of super symmetries, and $\cN = 4$ conformal supergravity. Amplitudes in these theories can be calculated explicitly using the accompanying {\sc Mathematica} package {\tt treeamps4dJAF}, which implements the algorithms described in this paper. When integrands are conjectured for new theories it will now be straightforward to test and evaluate them using the package. We include some discussion of the key functions from the package in this paper, and in the supporting files we provide the {\sc Mathematica} package along with example code and full documentation, and a lookup table with solutions up to 12 points NMHV and 9 points in all MHV sectors.

A natural question to ask is whether these techniques are applicable to the general $d$ scattering equations. It should be simple to map a full set of solutions at $n$ points from the 4D scattering equations to the general $d$ scattering equations for $d$ = 4, which would allow calculating amplitudes for the wider variety of theories supported by these equations in four dimensions. Solving the general $d$ equations directly using this algorithm is more difficult as they are not refined by helicity degree and all $(n-3)!$ solutions must be found. {\sc Mathematica}'s inbuilt algorithms can already solve the general $d$ equations numerically up to 9 points with a deterministic algorithm, so it would be necessary to improve on the efficiency of the Monte Carlo algorithm if it is to become a viable method for this problem.

{\tt treeamps4dJAF} provides only a basic implementation of the Monte Carlo equation solving algorithm, and there are a number of different ways that it could be improved. {\tt NSolveMonteCarlo} does not support parallel computation of different calls to {\tt FindRoot}, and does not use an optimal algorithm for deciding whether solutions are duplicate. The sampling method used where all initial points are taken from an approximated uniform distribution could be improved on. An updated algorithm with these changes in {\sc Mathematica} should have significantly better behaved time complexity. Re-writing the core solution finding algorithm in a different programming language better suited to low-level numerical computation such as {\tt C++} could also increase efficiency dramatically. These changes should allow solution finding in the 4D case for higher $n$ and $k$, and it is possible that they could allow solving the general $d$ equations by Monte Carlo algorithm to become viable. 

It is intriguing to find that the Cauchy distribution arises in the statistical analysis of the solutions to the 4D scattering equations, and especially to find that the parameter of the distribution is insensitive to changing $n$ and $k$. It would be interesting to explore how the functional form of this distribution is related to the physics of the equations; the statistical analysis of this paper is dependent on choice of GL(2) gauge, but it would be possible to do a similar analysis of GL(2) invariants.

A clear limitation of the 4D scattering equations is that they currently only support tree-level calculations, and an obvious future direction is to investigate calculation of loop-level integrands using these methods. Loop-level scattering equations exist in general dimensions\cite{Geyer:2017ela,Geyer:2018xwu}, and future work could implement numerical algorithms to evaluate loop-level integrands as sums over solutions to the general $d$ equations.

The 4D scattering equations are not restricted only to calculating amplitudes, they also cover form factors \cite{Brandhuber:2016xue} and possibly further structures in quantum field theory. They can also be used to calculate more physically relevant standard model amplitudes\cite{He:2016dol}. The tools we have provided in this paper should be directly relevant to calculation of form factors, and could play a role in finding new types of structures that can be supported on scattering equations.

\begin{center}
\textbf{Acknowledgements}
\end{center}

I would like to thank Simon Badger, Jake Bourjaily, Paul Heslop and Dan Rutter for useful discussions, Tim Whitbread for help with numerical algorithms, Themis Botsas for guidance on statistical methods, Theresa Abl for bug testing the package, the community at {\tt mathematica.stackexchange.com} for help with many {\sc Mathematica} related questions, and especially Arthur Lipstein, without whose excellent supervision I wouldn't have been able to complete this work. I have been funded for this work by EPSRC PhD scholarship EP/L504762/1. 

\appendix

\section{The {\sc Mathematica} Package {\tt treeamps4dJAF}}
\label{sec:mathematica}

{\tt treeamps4dJAF} provides a set of computational tools in {\sc Mathematica} for analytical and numerical calculation of amplitudes at tree-level. The package's most high-end functions are tailored for calculations in the 4D scattering equation formalism using the techniques derived in this paper. For example the package provides functionality for calculating MHV amplitudes analytically with specified external states in the supermultiplets of Yang-Mills and Einstein gravity theories, and graviton multiplets in conformal supergravity. Moving out of the MHV sector the package also provides functionality for solving the scattering equations numerically and using these solutions to calculate amplitudes. At a more low-level analysis of the code, many functions and a framework are provided for analytical computations in general in the spinner helicity formalism for amplitudes, including for example functions for dealing with antisymmetric brackets and the Schouten identity, and for evaluating expressions in terms of momentum twistors numerically. Each function is detailed in the documentation of the package, and can be accessed using {\tt Information} via {\tt ?Amplitude} or {\tt ??Amplitude}. Here we provide an overview of the package's key functions.

The packages reserves, clears and protects a number of default heads for usage with symbolic calculations. These can be accessed using {\tt GetHeads} and {\tt SetHeads}, and they are explained in \fref{fig:packageheads}.

\begin{figure}[h]
\centering
\begin{tabular}{|c|c|c|}
\hline
Name & Default Head & Mathematical Notation \\
\hline
{\tt "Angle"} & {\tt ang} & {\tt ang[i,j]} $= \ang{ij} = \det(\ket{i} \ket{j})$ \\
&&  {\tt ang[i,j,k,l]} $= \ang{ijkl} = \det(Z_iZ_jZ_kZ_l)$ \\
{\tt "Square"} & {\tt squ} & {\tt squ[i,j]} $= \squ{ij} = \det(\sket{i} \sket{j})$ \\
{\tt "Circle"} & {\tt cir} & {\tt cir[i,j]} $= (ij) = \det(\sigma_i \sigma_j)$ \\
{\tt "TwistorMu"} & {\tt mu} & {\tt mu[i,j]} $= \squ{\mu_i\mu_j} = \det(\mu_i \mu_j)$\\
{\tt "Lambda"} & {\tt lam} & {\tt lam[i]} $= \ket{i}$,\quad {\tt lam[i,a]} $= \ket{i}^a$\\
{\tt "LambdaTilde"} & {\tt lamt} & {\tt lamt[i]} $= \sbra{i}$,\quad {\tt lamt[i,a]} $= \sbra{i}^a$\\
{\tt "Worldsheet"} & {\tt s} & {\tt s[i,a]} $= \sigma_i^a$\\
{\tt "Grassmann"} & {\tt eta} & {\tt eta[i,A]} $= \eta_i^A$,\quad {\tt eta[A]} $= \eta^A$\\
\hline
\end{tabular}

    \caption{The default heads reserved by {\tt treeamps4dJAF} for analytical computations. See \sref{sec:review} for definitions and conventions.} 
    \label{fig:packageheads}
\end{figure}

\subsection{Calculating Amplitudes}
The most high-level functions of {\tt treeamps4dJAF} are used for calculation and readable display  of analytical and numerical amplitudes.\\[-.2cm]

\noindent\fun{Amplitude} is the package's main function. It is used to calculate tree-level analytical MHV amplitudes and numerical amplitudes in all MHV sectors by solving the 4D scattering equations. Analytical output of \fun{Amplitude} will be a function of the external momenta in terms angle and square brackets and Grassmann $\eta$ variables; see \fref{fig:packageheads} for details. The input for \fun{Amplitude} is overloaded in a number of different ways. Analytical amplitudes are specified in the form;
\begin{itemize} 
\setlength\itemsep{-.1cm}
	\item \fun{Amplitude}[\var{theory}, \var{N}, \var{states}]
	\item \fun{Amplitude}[\var{theory}, \var{N}, \var{n}, \var{L}]
\end{itemize}
Numerical amplitudes can take any of the following forms;
\begin{itemize}
\setlength\itemsep{-.1cm}
	\item \fun{Amplitude}[\var{theory}, \var{N}, \var{momenta}, \var{states}]
	\item \fun{Amplitude}[\var{theory}, \var{N}, \var{momenta}, \var{n}, \var{L}]
	\item \fun{Amplitude}[\var{theory}, \var{N}, \var{momenta}, \var{L}$_{solutions}$, \var{solutions}, \var{states}]
	\item \fun{Amplitude}[\var{theory}, \var{N}, \var{momenta}, \var{L}$_{solutions}$, \var{solutions}, \var{L}]
\end{itemize}
\var{n} and \var{L} specifications can be used to calculate Grassmann superamplitudes, and providing external states will perform the Grassmann integrations to give a component amplitude. If a set of numerical \var{momenta} matching the state specification are entered, \fun{Amplitude} will attempt to solve the relevant scattering equations, and return either a numerical value or a superamplitude Grassmann expansion with numerical coefficients. On entering a set of numerical \var{momenta} along with left set \var{L}$_{solutions}$ and a full set of solutions \var{solutions} which match the number of points and MHV degree of the specified states, \fun{Amplitude} will sum the solutions over the relevant integrand and return the numerical amplitude.
	
\mathematicaaa{Amplitude["EG",0,"h-","h-","h+"]}{$\frac{\tt ang[1, 2]^6}{\tt ang[1, 3]^2 ang[2, 3]^2}$}
{Amplitude["YM", 1, 3, \{1, 2\}]}{$\frac{\tt ang[1, 2]^2 (-ang[1, 2] eta[2, 1] ** eta[1, 1] - ang[1, 3] eta[3, 1] ** eta[1, 1] -
  ang[2, 3] eta[3, 1] ** eta[2, 1])}{\tt ang[1, 3] ang[2, 3]}$}
{Amplitude["EG", 0, RandomMomenta4D[8], \{1, 3, 5, 7\}]}{11.8556 - 30.6627 I}

\noindent \fun{AmplitudesDisplay}[\var{expr}] displays all of the package's heads (as explained in \fref{fig:packageheads}) in a simplified format for readability.

\mathematicaa{Amplitude["YM",0,"g-","g-","g+","g+"]//AmplitudesDisplay}{
$-\frac{\tt\ang{12}^3}{\tt\ang{23} \ang{34}\ang{14}}$}
{Amplitude["YM", 2, 3, \{1\}] // AmplitudesDisplay}
{$\frac{[23]([23]\eta_1^1 + [12]\eta_3^1 -[13]\eta_2^1)([23]\eta_1^2 + [12]\eta_3^2 -[13]\eta_2^2)}{\tt [12] [13]}$}

\noindent\fun{ListTheories}[] returns a list of all theories with number of supersymmetries currently supported by {\tt treeamps4dJAF}.

\mathematica{ListTheories[]}{
\{\{"YM", 0\}, \{"YM", 1\}, \{"YM", 2\}, \{"YM", 3\},\{"YM", 4\}, \{"EG", 
  0\}, \{"EG", 1\}, \{"EG", 2\}, \{"EG", 3\}, \{"EG", 4\}, \{"EG", 5\}, \{"EG", 
  6\}, \{"EG", 7\}, \{"EG", 8\}, \{"CG", 4\}\}}

\noindent\fun{ListStates}[\var{theory}, \var{N}] returns a list of the specifications for all possible external states in \var{theory} with \var{N} supersymmetries, possibly with R symmetry indices.

\mathematica{ListStates["YM",2]}{\{"g-", \{"\textPsi-", A\}, "\textPhi-", "\textPhi+", \{"\textPsi+", A\}, "g+"\} }

\subsection{Solving the 4D Scattering Equations}
The package's core functions for working with the 4D scattering equations are\\[-.2cm]

\noindent \fun{ScatteringEquations4D}[\var{n}, \var{L}, \var{gaugefix}, \var{delete}] returns the scattering equations at \var{n} points with left set \var{L}, with worldsheet variables specified by \var{gaugefix} fixed to the identity, and with equations given by \var{delete} removed.
\noindent  \fun{ScatteringEquations4D}[\var{momenta}, \var{L}, \var{gaugefix}, \var{delete}] returns the equations as a function of worldsheet minors only, for the specified numerical \var{momenta}. If  \var{gaugefix} and \var{delete} are not entered, \fun{ScatteringEquations4D} returns the scattering equations at \var{n} points with left set \var{L}  in terms of worldsheet minors.

\mathematicaa{ScatteringEquations4D[3, \{1, 2\}, \{1, 2\}]}{
${\tt\{lam[3, 1] - \frac{lam[2, 1]}{s[3, 1]} + \frac{lam[1, 1]}{s[3, 2]}, 
 lam[3, 2] - \frac{lam[2, 2]}{s[3, 1]} + \frac{lam[1, 2]}{s[3, 2]}\}}$}
 {ScatteringEquations4D[4, \{1, 2\}] // AmplitudesDisplay}
 {${\tt\{
 \sbra{1}-\frac{\sbra{3}}{(13)}-\frac{\sbra{4}}{(14)},
 \sbra{2}-\frac{\sbra{3}}{(23)}-\frac{\sbra{14}}{(24)},
 \ket{3}-\frac{\ket{1}}{(31)}-\frac{\ket{2}}{(32)},
 \ket{4}-\frac{\ket{1}}{(41)}-\frac{\ket{2}}{(42)}
 \}}$}

\noindent\fun{SolveScatteringEquations4D}[\var{n}, \var{L}, \var{gaugefix}] returns the analytical MHV and \MHVB solutions to the scattering equations derived in \sref{sec:analytical} at \var{n} points with left set \var{L} and particles specified by \var{gaugefix} gauge-fixed to the identity.

\mathematica{SolveScatteringEquations4D[4, \{1, 2\}]}{
${\tt \{s[1, 1] \rightarrow 1, s[1, 2] \rightarrow 0, s[2, 1] \rightarrow 0, s[2, 2] \rightarrow 1, 
 s[3, 1] \rightarrow \frac{ang[1, 2]}{ang[1, 3]}, }$ ${\tt
 s[3, 2] \rightarrow \frac{ang[1, 2]}{ang[2, 3]}, 
 s[4, 1] \rightarrow \frac{ang[1, 2]}{ang[1, 4]}, s[4, 2] \rightarrow \frac{ang[1, 2]}{ang[2, 4]}\}}$}

\noindent\fun{NSolveMonteCarlo}[\var{equations}, \var{variables}, \var{inputsolutions}] attempts to find numerical approximations to the solutions of the system \var{equations} for the \var{variables} by the Monte Carlo algorithm outlined in \sref{sec:montecarlo}. \var{inputsols} is an optional argument for providing a list of previously known solutions to the equations. There are a number of different options for specifying details of the algorithm. 
{\tt NumSolutions}, {\tt MaxIterations} and {\tt MaxTime} provide different stopping conditions, and {\tt Distribution} sets the sampling distribution for initial points. Other options can be checked in the documentation. Calling {\tt Abort} ({\tt Alt + .}) during the calculation will return any solutions currently found.
	
\mathematica{NSolveMonteCarlo[\{x$^{\tt x}$ + x$^{\tt 4-x}$ - 4\}, \{x\}, NumSolutions~$\rightarrow$~5]}{
\{\{-0.864475 - 0.526651 I\}, \{1.39997\}, \{0.00678977 + 
   0.95836~I\}, \{0.00678977 - 0.95836~I\},\{-0.864475 + 0.526651 I\}\}
}

\noindent\fun{NSolveScatteringEquations4D}[\var{momenta}, \var{L}, \var{gaugefix}, \var{delete}, \var{inputsolutions}] attempts to find numerical approximations for all of the ${\rm \ang{\begin{smallmatrix}{\tt Length}[\var{momenta}]-3\\{\tt Length}[\var{L}]-2\end{smallmatrix}}}$ solutions to the scattering equations defined by numerical \var{momenta} and left set \var{L} via the \fun{NSolveMonteCarlo} algorithm. Optional argument \var{gaugefix} specified which worldsheet variables are gauge-fixed to the identity, and \var{delete} specifies which equations are removed. The initial random points are sampled from the distributions described in \sref{sec:montecarlo}. Any previously known solutions can be provided via \var{inputsolutions}.

\mathematica{NSolveScatteringEquations4D[RandomMomenta4D[4], \{1, 2\}]}{
\{\{\{1, 0\}, \{0, 
   1\}, \{1.14552 - 1.31495 I, -1.08314 - 0.751286~I\}, \{0.191178 + 
    0.920906 I, -0.461597 + 0.613013 I\}\}\}}

\subsection{Applying Momenta and Solutions to Expressions}
{\tt treeamps4DJAF} provides a number of functions for evaluating analytical expressions on specific numerical external data.\\[-.2cm]

\noindent\fun{RandomMomenta4D}[\var{n}, \var{w}] samples a set of complex numbers with rational real and imaginary parts in the interval [-\var{w}, \var{w}], and returns \var{n} point complexified momenta satisfying momentum conservation. Each element is in the form $\{\ket{i},\sbra{i}\}$, and momentum conservation is solved by setting $\sket{1} = \sum_{i = 3}^n\frac{\ang{2i}}{\ang{12}}\sket{i}$ and $\sket{2	} = \sum_{i = 3}^n\frac{\ang{1i}}{\ang{21}}\sket{i}$.\\[-.2cm]

   
\noindent\fun{ApplyMomenta}[\var{expression}, \var{momenta}] replaces the package's heads for spinor helicity variables with the \var{momenta} in the given \var{expression}.

\mathematica{ApplyMomenta[ang[1, 2] lam[3], RandomMomenta4D[6]]}{
\{6141/250000 - (49771 I)/1000000, 
 4071641249/125000000000 + (22520160139 I)/1000000000000\}
}

\noindent \fun{Integrand} returns a theory-dependent tree-level integrand which can be summed over solutions to the scattering equations to return an amplitude. The output of \fun{Integrand} will be an analytical expression which is a function of the worldsheet and external momenta, and Grassmann $\eta$ variables.
The input for \fun{Integrand} is overloaded in different ways, for example;
\begin{itemize}
	\item \fun{Integrand}[\var{theory}, \var{N}, \var{states}]
	\item \fun{Integrand}[\var{theory}, \var{N}, \var{n}, \var{L}]
\end{itemize}
The specifications for the states work the same as for \fun{Amplitude}, which inherits its specifications from \fun{Integrand}.

\mathematicaa{Integrand["EG", 0, 4, \{1, 2\}]}{${\tt \frac{ang[1, 2] squ[3, 4]}{cir[1, 2] cir[3, 4]}}$}
{Integrand["CG", 4, "h-", "h-", "h+", "h+"]}{${\tt \frac{ang[1, 2]^2 squ[3, 4]^2}{cir[1, 2]^2 cir[3, 4]^2}}$}

\noindent\fun{ApplyScatteringSolutions}[\var{expression}, \var{momenta}, \var{L}, \var{solutions}] applies numerical \var{momenta} to \var{expression}, and then sums \var{expression} and the relevant Jacobian over the given solutions to the scattering equations, for the number of points specified by the length of \var{momenta} and left set \var{L}.

\mathematica{mom = RandomMomenta4D[6];

sols = NSolveScatteringEquations4D[mom, \{1, 2, 3\}];

ApplyScatteringSolutions[cir[1,2]cir[3,4]cir[5,6], 
mom, \{1,~2,~3\}, sols]}
 {-23.1883 + 27.2079 I}

\subsection{Grassmann Integration and Swapping Left Set Algorithms}

\fun{DetGIntegrate}[\var{expr}, \var{measure}, \var{space}] implements the algorithm from \sref{sec:grassmann} for efficient Grassmann integration.

\mathematica{DetGIntegrate[(a eta[1] + b eta[2]) ** (c eta [1] + 
    d eta[2]), \{\{1\}, \{2\}\}, \{\{1\}, \{2\}\}]}{-b c + a d} 
    
\noindent \fun{GIntegrateStates}[\var{theory}, \var{N}, \var{expression}, \var{states}] selects the relevant Grassmann $\eta$ variables associated with the \var{states} for \var{theory} with \var{N} supersymmetries, and uses these as the measure to integrate \var{expression}.

\mathematica{GIntegrateStates["YM", 4, MHVBarGrassmannFactor[4, 3], "g-", "g+", "g+"]}{squ[2, 3]$^4$}

\noindent\fun{SwapParticles} [\var{integrand}, \var{n}, \var{L}, \var{l0}, \var{r0}] applies the worldsheet transformation specified in \sref{sec:perms} to the \var{n} point \var{integrand} with left set \var{L}, swapping \var{l0} and \var{r0} in between the left set and the right set.\\[-.2cm]

\noindent \fun{ChangeLeftSet}[\var{integrand}, \var{n}, \var{Lold}, \var{Lnew}] applies \fun{SwapParticles} repeatedly until the \var{integrand} supported on \var{Lold} is supported on \var{Lnew}

\mathematica{ChangeLeftSet[1/(cir[1, 2] cir[2, 3] cir[3, 1]), 3, \{1, 2\}, \{1, 3\}]}{${\tt-\frac{1}{cir[1, 2] cir[1, 3] cir[2, 3]^5}}$}

\subsection{File Input-Output for Sets of Solutions}
Solving the 4D scattering equations using \fun{NSolveScatteringEquations4D} is stochastic, and can take a long time for higher points and MHV degrees. {\tt treeamps4dJAF} provides functions for storing and reading set of Solutions to and from a CSV file, along with a lookup table of solutions in {\tt SolutionLookupTable.csv} which can be used without solving the equations. \\[-.2cm]

\noindent\fun{ReadSolution}[\var{filename}, \var{n}, \var{L}, \var{solutionnumber}] looks for a solution specified by \var{n} and \var{L} in \var{filename}{\tt .csv}, and returns the relevant data if found.

\mathematica{ReadSolution["SolutionLookupTable", 4, \{1, 2\}]}{\{\{\{1, 0\}, \{0, 
   1\}, \{1.14552 - 1.31495 I, -1.08314 - 0.751286 I\}, \{0.191178 + 
    0.920906 I, -0.461597 + 0.613013 I\}\}\}}

\noindent Importing large CSV files directly into {\sc Mathematica} can be very slow. \\[-.2cm]

\noindent \fun{ReadSolutionData}[\var{filename}] opens \var{filename}.csv and returns the data for each row of \var{filename}{\tt .csv} in the form {\tt\{number of points, left set, gauge-fixed rows\}} without actually loading the solutions or momenta into the RAM.\\[-.2cm]

\noindent\fun{AppendSolution}[\var{filename}, \var{momenta}, \var{L}, \var{solutions}] opens \var{filename}{\tt .csv} and appends at the end of it a new row containing the data {\tt\{Length[\var{momenta}], L,\linebreak GetGaugeFix[solutions], momenta, solutions\}}, with the gauge-fixed columns removed from \var{solutions}.\\[-.2cm]

\noindent \fun{SolveAppendSolution}[\var{filename}, \var{momenta}, \var{L}, \var{inputsolutions}] attempts to find numerical solutions to the scattering equations and then append them to \var{filename}{\tt .csv}.

\section{Analytical Details of the 4D Scattering Equations}
\label{sec:analappen}
In this appendix we provide detailed proofs of the analytical results from \sref{sec:analytical}, along with some further general results for the 4D scattering equations.

\subsection{Recovering the General $d$ Scattering Equations for $d$ = 4}
\label{sec:generald}
Solutions to the 4D equations are grouped into sets for the different N$^{k-2}$MHV sectors. The general $d$ equations depend only on $n$ and hence do not encode this grouping. As the 4D scattering equations contain more information than the general~$d$ scattering equations we have 4D~$\Rightarrow$~general~$d$ (for $d$ = 4), which we prove in this section. A different argument is given in \cite{Dolan:2014ega}. The proof also provides an explicit method for finding solutions to the general~$d$ equations using those from the 4D specific case. Integrands for the general $d$ equations for $d = 4$ can also be mapped to integrands for the 4D equations \cite{Roehrig:2017wvh}. 

First we prove a lemma which holds for the general $d$ equations. Define a world-sheet dependent momentum $P(s) :=  \sum_{j\in N}\frac{k_i}{s - s_i}$. Then we prove that $P(s)^2~=~0 \Leftrightarrow$ the general $d$ equations. Note that there are no second order poles in $P(s)^2$ as all of the external momenta $k_i$ are null.

\begin{equation}
\begin{split}
P(s)^2  =& \sum_{i,j\in \mathscr{N}\atop i \neq j}\frac{k_i\cdot k_j}{(s - s_i)(s - s_j)} =2 \sum_{i \in \mathscr{N}}\frac{1}{s - s_i}\left(\sum_{j\in \mathscr{N}\atop j \neq i}\frac{k_i\cdot k_j}{s_i - s_j}\right),
\end{split}
\end{equation}
\noindent where we use partial fractions and relabel indices to arrive at the result. $P(s)$ is now written explicitly as a sum of its poles, and hence can only be zero if all of its residues are zero. Then we conclude that $P(s)^2~=~0 \Leftrightarrow \sum_{j\in \mathscr{N}}\frac{k_i \cdot k_j}{s_i - s_j}=0$.

To relate to the 4D equations, construct a new explicitly null world-sheet dependent momentum in terms of two world-sheet dependent spinors $\ket{\lambda(s)} := \sum_{r\in R} \frac{t_r \ket{r}}{s - s_r}$ and $\sket{\lambda(s)} := \sum_{l\in L} \frac{t_l \sket{l}}{s - s_l}$. 
\begin{equation}
\begin{split}
&\sket{\lambda(s)}\bra{\lambda(s)} = \left(\sum_{r\in R} \frac{t_r \sket{r}}{s - s_r}\right) \left(\sum_{l\in L} \frac{t_l \bra{l}}{s - s_l}\right) = \sum_{r\in R\atop l\in L} \frac{\sket{r} \bra{l}}{(rl)}\left(\frac{1}{s - s_l}-\frac{1}{s - s_r}\right)\\
&\hspace{.5cm}= \sum_{r\in R} \frac{\sket{r}}{s - s_r}\left(\sum_{l\in L}\frac{\bra{l}}{(rl)}\right) + 
\sum_{l\in L}\left(\sum_{r\in R}\frac{\sket{r}}{(lr)}\right) \frac{\bra{l}}{s - s_l} 
\\&\hspace{.5cm}=  \sum_{r\in R} \frac{\sket{r}\bra{r}}{s - s_r} + 
\sum_{l\in L}\frac{\sket{l}\bra{l}}{s - s_l} = P(s),
\end{split}
\end{equation}
\noindent where we use the same steps as in the previous calculation, and in the second last equality we use the 4D scattering equations and $k_i = \sket{i}\bra{i}$. Given that $\sket{\lambda(s)}\bra{\lambda(s)}$ is explicitly constructed as a null vector, we then see that the 4D scattering equations imply that $P(s)^2 = 0$, and hence by the lemma they imply the general $d$ scattering equations.

This proof also provides an explicit mapping from a solution to the 4D scattering equations to a solution to the general $d$ equations. We map a point in the solution space of the 4D equations to a point in the $n$-punctured Riemann sphere by writing each column as $\sigma_i= t_i^{-1}\left(\begin{smallmatrix}1 \\ s_i \end{smallmatrix}\right)$, and keeping only the $s$ variables and not the scales $t$. Suppose that we have some solution to the 4D scattering equations with the gauge-fixing that the first two particles are equal to the identity matrix. Under this mapping the vector $ \left(\begin{smallmatrix}1 \\0 \end{smallmatrix}\right)$ maps to the point at infinity, and we fix the remaining gauge redundancy by dividing through by $s_3$ to arrive at

\begin{equation}
\begin{split}
\label{eqn:gentofour}
\sigma_{\rm 4D} &= 
\begin{pmatrix}
1 & 0 & \sigma_3^1& ... & \sigma_n^1\\
0 & 1 & \sigma_3^2& ... & \sigma_n^2
\end{pmatrix}
\rightarrow s_{{\rm general }\, d, d = 4} = \begin{pmatrix}\infty & 0 & 1 & \frac{\sigma_4^1}{\sigma_4^2}\frac{\sigma_3^2}{\sigma_3^1}& ... & \frac{\sigma_n^1}{\sigma_n^2}\frac{\sigma_3^2}{\sigma_3^1}\end{pmatrix}.
\end{split}
\end{equation}

From this analysis we see that reconstructing a full solution to the 4D equations in terms of the $s$ and $t$ variables is not direct given a $d$ = 4 solution to the general $d$ equations, and it would be interesting to understand how the $t$ variables can be specified in this case.

MHV and \MHVB solutions to the general $d$ equations for $d = 4$ were derived in \cite{Weinzierl:2014vwa}. Using the mapping~\ref{eqn:gentofour}, we can see explicitly how the MHV solutions to the 4D equations derived in \sref{sec:analytical} map onto those of the general $d$ equations

\begin{equation}
\begin{split}
\sigma_{\text{MHV}} = \left(\begin{matrix}
1 & 0 & \frac{\ang{12}}{\ang{31}} &...& \frac{\ang{12}}{\ang{n1}}\\
0 & 1 & \frac{\ang{12}}{\ang{32}} &...& \frac{\ang{12}}{\ang{n2}}
\end{matrix}\right)
\rightarrow s_{\text{MHV}} &= \begin{pmatrix}\infty & 0 & 1 & \frac{\ang{41}\ang{32}}{\ang{31}\ang{42}}&...&\frac{\ang{n1}\ang{32}}{\ang{31}\ang{n2}}\end{pmatrix},
\end{split}
\end{equation}
in agreement with equation (49) of \cite{Weinzierl:2014vwa}, up to choice of SL(2) fixing.

\subsection{Symmetries, Little Group Scaling and Grassmanians}
\label{sec:syms}
The scattering equations have a GL(2) symmetry which can be realised in different ways in terms of a worldsheet redefinition, or a worldsheet redefinition with a corresponding little group rescaling. The worldsheet GL(2) symmmetry in (\ref{eqn:GL2}) is a combination of the standard SL(2) symmetry of global conformal transformations in the string worldsheet $s$ variables, and a GL(1) transformation corresponding to a rescaling of the worldsheet $t$ variables. Any function $f(\sigma)$ which is integrated against the scattering equation delta functions must transform as $f(\sigma)\rightarrow f(\sigma)(\det G)^{n-2k}$ under (\ref{eqn:GL2}) to balance out the transformation of the measure. All of the integrands for the theories considered in \sref{sec:review} satisfy this transformation law, as enforced by their underlying 4D ambitwistor string models \cite{Geyer:2014fka}.


Before considering joint worldsheet and little group transformations, we first analyse the little group scaling of amplitudes supported on the scattering equations. First consider a general amplitude $\cA_{n,L}$ with some arbitrary integrand $f(\sigma, \ket{i}, \sket{i})$,

$$
\cA_{n,L} := \int \frac{d^{2\times n}\sigma}{GL(2)}  \diracd{2\times n}{SE^n_L}f(\sigma, \ket{i}, \sket{i}).
$$

\noindent Perform a different little group scaling for each particle, such that $\ket{i} \rightarrow \alpha_i \ket{i}$, $\sket{i} \rightarrow \alpha_i^{-1} \sket{i}$. Then we see that

\begin{equation}
\begin{split}
\cA_{n,L}(\alpha_i \ket{i}, \alpha_i^{-1} \sket{i}) =  \int \frac{d^{2\times n}\sigma}{GL(2)} 
\prod_{l \in L} \diracd{2}{\frac{|l]}{\alpha_l} - \sum_{r\in R}\frac{|r]}{\alpha_r(lr)}}\prod_{r \in R}&\diracd{2}{\alpha_r\ket{r} - \sum_{l\in L}\frac{\alpha_l\ket{l}}{(rl)}}\\
&f(\sigma, \alpha_i \ket{i}, \alpha_i^{-1} \sket{i}).
\end{split}
\end{equation}

\noindent Define new worldsheet coordinates such that $\sigma_l' := \alpha_l^{-1} \sigma_l$ for $l \in L$, and $\sigma_r' := \alpha_r \sigma_r$ for $r \in R$. Change variables and rename back to $\sigma$, picking up factors of the $\alpha_i$ from the delta functions and the measure to arrive at

\begin{equation}
\begin{split}
\cA_{n,L}(\alpha_i \ket{i}, \alpha_i^{-1} \sket{i}) &= \int \frac{d^{2\times n}\sigma}{GL(2)} \diracd{2\times n}{SE^n_L} \prod_{l\in L}\alpha_l^4\prod_{r\in R}\alpha_r^{-4}\,\, f(\alpha_l \sigma_l,\alpha_r^{-1} \sigma_r, \alpha_i \ket{i}, \alpha_i^{-1} \sket{i}).
\end{split}
\end{equation}

Now consider how the integrand for Yang-Mills theory with $\cN$ supersymmetries as defined in \sref{sec:review} scales under the little group. Under this little group transformation the Grassmann variables transform as $\eta_i \rightarrow \alpha_i^{-1} \eta_i$, and the Grassmann delta functions transform in a similar way to the scattering equation delta functions. The transformation of the Parke-Taylor factor cancels out that of the measure, and the integrand becomes 

\begin{equation}
\begin{split}
f_{\text{sYM}}(\alpha_l \sigma_l,\alpha_r^{-1} \sigma_r, \alpha_i \ket{i}, \alpha_i^{-1} \sket{i}, \alpha_i^{-1} \eta_i) =  \prod_{l\in L}\alpha_l^{-2-\cN}\prod_{r\in R}\alpha_r^{2} \,\,f_{\text{sYM}}(\sigma,\ket{i}, \sket{i},\eta_i).
\end{split}
\end{equation}

\noindent From this we find the standard little group scaling for negative and positive helicity superfields of a Yang-Mills superamplitude, that 
\begin{equation}
\begin{split}
\cA_{n,L}(\alpha_i \ket{i}, \alpha_i^{-1} \sket{i}, \alpha_i^{-1} \eta_i) = \prod_{l\in L}\alpha_l^{2-\cN}\prod_{r\in R}\alpha_r^{-2}\cA_{n,L}(\ket{i}, \sket{i},\eta_i).
\end{split}
\end{equation}

\noindent A similar analysis produces the required scaling for gravity amplitudes.

Combining these two types of transformations we find a symmetry which acts with a standard GL(2) transformation such that $\sigma \rightarrow G \sigma$, and hence the minors of the $\sigma$ matrix transform as $(ij) \rightarrow \det G(ij)$. Simultaneously performing an inhomogeneous little group scaling such that  $|l] \rightarrow \alpha |l], \ket{l} \rightarrow \alpha^{-1} \ket{l}$ for $l \in L$ and $|r] \rightarrow \beta |r], \ket{r} \rightarrow \beta^{-1} \ket{r}$ for $r \in R$, the scattering equations become 

\begin{equation}
\begin{split}
\diracd{2\times n}{SE^n_L} &= \prod_{l\in L}\alpha^{-2}\diracd{2}{|l] - \frac{\beta}{\alpha\det G}\sum_{r\in R}\frac{|r]}{(lr)}}\prod_{r\in R}\beta^{2}\diracd{2}{
\ket{r} -\frac{\beta}{\alpha\det G} \sum_{l\in L}\frac{ \ket{l}}{(rl)}} \\
&= \alpha^{2n-4k}(\det G)^{2n-2k}\diracd{2\times n}{SE^n_L},
\end{split}
\end{equation}
\noindent where in the last equation we choose $\beta = \alpha \det G$ to keep the equations invariant. The measure and delta functions combined transform such that

\begin{equation}
\begin{split}
&\hspace{-.5cm}\int \frac{d^{2\times n}\sigma}{GL(2)} \diracd{2\times n}{SE^n_L}  f(\sigma, \ket{i}, \sket{i})\rightarrow
\int \frac{d^{2\times n}\sigma}{GL(2)} \diracd{2\times n}{SE^n_L}(\det G)^{3n-2k}\alpha^{2n-4k}\\
&\hspace{7cm}  f(G \sigma, \alpha \ket{l}, \alpha^{-1} \sket{l},\alpha\det G \ket{r}, (\alpha\det G)^{-1} \sket{r}).
\end{split}
\end{equation}

We have seen how the little group transformation of the amplitude $\cA$ comes from little group covariance of the integrand $f$. Any $f$ which integrates to an amplitude must transform covariantly under the little group, and hence we need that $f$ transforms as a scaling transformation for any combined little group transformation and worldsheet rescaling. We can then conclude that for any $f$ which describes an amplitude there must exist some $x$ and $y$ real numbers such that 

$$
f(G \sigma, \alpha \ket{l}, \alpha^{-1} \sket{l},\alpha\det G \ket{r}, (\alpha\det G)^{-1} \sket{r}) = (\det G)^{x}\alpha^y f(\sigma,\ket{i}, \sket{i}).
$$
Given $f$ which transforms in this way, we can choose the little group scaling $\alpha$ to such that $(\det G)^{3n-2k+x}\alpha^{2n-4k+y}=1$, and hence this transformation is a symmetry for any amplitude supported on the scattering equations. This GL(2) invariance is the GL(2) invariance of the Grassmannian Gr(2,~$n$), and hence in this sense we can think of the solutions to the scattering equations as living in Gr(2,~$n$).

\subsection{Deleting Equations and Momentum Conservation}
\label{sec:gfix}

In this appendix we demonstrate how to remove four equations to give a momentum conserving delta function. There are three possible cases; we either consider the equations for two particles in the left set, or for two in the right set, or for one in each. First consider the case of two particles in the left set. Without loss of generality, label these particles to be 1 and 2. Defining the delta functions for these particles to be $\Delta_{1,2}$ we see that
\begin{equation}
\begin{split}
\Delta_{1,2} &:=\delta^2\left(|1] - \sum_{r\in R}\frac{|r]}{(1r)}\right)\delta^2\left(|2] - \sum_{r\in R}\frac{|r]}{(2r)}\right)\\
&= \ang{12}^4\delta^2\left(|1]\ang{12} - \sum_{r\in R}\frac{|r]\ang{12}}{(1r)}\right)\delta^2\left(|2]\ang{21} - \sum_{r\in R}\frac{|r]\ang{21}}{(2r)}\right).
\end{split}
\end{equation}

Now consider a general right set equation and contract with first with $\bra{2}$, and separately with $\bra{1}$, to find that

\begin{equation}
\begin{split}
\ket{r} = \sum_{l\in L}\frac{ \ket{l}}{(rl)} 
\implies & \frac{\ang{12}}{(1r)} = -\ang{r2} -\sum_{l \in L\atop l \neq 1,2} \frac{\ang{l2}}{(lr)} \\
 & \frac{\ang{21}}{(2r)} = -\ang{r1} -\sum_{l \in L\atop  l \neq 1,2} \frac{\ang{l1}}{(lr)}.
\end{split}
\end{equation}
Substituting into $\Delta_{1,2}$, we see that

\begin{equation}
\begin{split}
\Delta_{1,2} &=
\ang{12}^4\delta^2\left(\left(|1]\bra{1} +\sum_{r\in R}|r]\bra{r}\right)\ket{2} + \sum_{r\in R ,\,l \in L\atop  l \neq 1,2}\frac{|r]\ang{l2}}{(lr)}\right)\\
&\hspace{2cm} \diracd{2}{\left(|2]\bra{2} +\sum_{r\in R}|r]\bra{r}\right)\ket{1} + \sum_{r\in R, \,l \in L\atop  l \neq 1,2}\frac{|r]\ang{l1}}{(lr)}}.
\end{split}
\end{equation}
We then use the remaining left set equations to solve the sum over $r$ in the last term in each delta function, arriving at

\begin{equation}
\begin{split}
\Delta_{1,2}
= \ang{12}^4\diracd{2}{\bigg(\sum_{n\in \mathscr{N}}|n]\bra{n}\bigg)\ket{2}} \diracd{2}{\bigg(\sum_{n\in \mathscr{N}}|n]\bra{n}\bigg)\ket{1}}
= \ang{12}^2\diracd{4}{P}.
\end{split}
\end{equation}

We would now say that we have `deleted equations 1 and 2', and the remaining $2n-4$ equations now give a well-specified system. Note the Jacobian $\ang{12}^2$ for this calculation. The calculation for deleting two equations in the right set goes by the same steps, and labelling the two particles in the right set to be $r_1$ and $r_2$ we find that

\begin{equation}
\begin{split}
\diracd{2}{\ket{r_1} - \sum_{l\in L}\frac{ \ket{l}}{(r_1l)}} 
\diracd{2}{\ket{r_2} - \sum_{l\in L}\frac{ \ket{l}}{(r_2l)}}  
= \squ{r_1r_2}^2\diracd{4}{P}.
\end{split}
\end{equation}

There is one remaining choice; we could delete one equation from the left set and one equation from the right set. Choosing equations in this way does not produce a momentum-conservation delta function, and hence does not result in a solvable system for spinors satisfying momentum conservation. Label the left set particle as $1\in L$, and the right set particle as $n\in R$. Then following the same analysis as above, we arrive at 

\begin{equation}
\begin{split}
\diracd{2}{\sket{1} - \sum_{r\in R}\frac{ \ket{r}}{(1r)}} 
&\diracd{2}{\ket{n} - \sum_{l\in L}\frac{ \ket{l}}{(nl)}}  \\
&= \ang{1n}^2\squ{1n}^2\diracd{2}{\sket{1}\ang{1n}+\sum_{i = 2}^{n-1}\sket{i}\ang{in}} \diracd{2}{\sum_{i = 2}^{n-1}\squ{1i}\bra{i}+\squ{1n}\bra{n}}.
\end{split}
\end{equation}
These equations looks deceptively similar to momentum conservation. We prove here that they are in fact not the same, and give a constraint corresponding to non-generic kinematics.

 Solving the first equation as $\sket{1} = \frac{1}{\ang{n1}}\sum_{i = 2}^{n-1}\sket{i}\ang{in}$ and substituting into the second, we find that these delta functions imply that

\begin{equation}
\begin{split}
&\ket{n}\sum_{i = 2}^{n-1} P_n\cdot P_i  +\sum_{i,j = 2\atop j\neq i}^{n-1}\ket{i}\squ{ij}\ang{jn}  
= \ket{n} \bigg(\sum_{i = 2}^{n-1} P_i \bigg)^2 = 0,
\end{split}
\end{equation}
where between the first two equalities we split the second sum into two terms, relabel the indices use a Schouten identity. Hence to keep consistency with these equations we must set either $\ket{n}$ or the Mandelstam invariant $\bigg(\sum_{i = 2}^{n-1} P_i \bigg)^2$ to zero, and neither of these choices correspond to generic kinematics. From this we see that it is not possible to delete one equation from each set.

\subsection{Permutations and Choice of Left Set}
\label{sec:perms}
A given $n$ point N$^{k-2}$MHV amplitude will be supported on scattering equations with a specified left set $L$. In this section we show how solutions to the scattering equations for one left set can be used to calculate amplitudes with the same MHV degree that are supported on different left set $L'$. This mapping of different left sets is important computationally as it will allow us to solve the equations only once in each MHV sector and then calculate all amplitudes of this MHV degree for the specified numerical momenta. The explicit worldsheet transformation swapping particles $l_0 \in L$ and $r_0 \in R$ in between the sets is given in \eref{eqn:permtrans}. Under this transformation, we find that the scattering equations transform as
\begin{align}
	E_l \quad &\longrightarrow \quad E_l-\frac{(l_0r_0)}{(lr_0)} E_{l_0}\notag
	&\qquad E_{l_0} \quad &\longrightarrow \quad {(l_0r_0)} E_{r_0}\notag\\
	E_r \quad &\longrightarrow \quad E_r - \frac{(r_0 l_0)}{(rl_0)} E_{r_0}\notag
		&\qquad E_{r_0} \quad &\longrightarrow \quad (l_0r_0) E_{l_0},
\end{align}
where we used a Schouten identity in the worldsheet variables for the $l$ and $r$ equations. The second terms in the $l$ and $r$ equations are zero on support of the $l_0$ and $r_0$ equations, and hence we see that the scattering equations remain the same up to changing the particles $l_0$ and $r_0$ between the left and right set. The delta functions pick up an overall factor of $(l_0 r_0)^{-4}$. Calculating the transformation of the measure, we find that the scattering equation integral transforms as in \eref{eqn:scattinttrans}.
 
 If we now also perform a permutation of the external data on the same legs $l_0$ and $r_0$ and look at the transformation properties of the integrands, we can understand how the amplitude does not depend on a choice of left set for maximally supersymmetric theories and for $\cN = 4$ conformal supergravity. This transformation can also be used to understand permutation invariance under swapping between the right and left sets for $\cN = 8$ supergravity. Showing permutation invariance of gravity amplitudes under a swapping two legs which carry the same helicity superfield is straightforward, and simply requires renaming the worldsheet variables on the permuted legs.
 
 Let us now look at the form of the integrand for Yang-Mills theories. Under this transformation, we find that

\begin{equation}
\begin{split}
&\int \frac{d^{2\times n}\sigma}{GL(2)} \diracd{2\times n}{SE^n_L}  \frac{\prod_{l\in L} \diracd{\cN}{\eta_l - \sum_{r\in R}\frac{\eta_r}{(lr)}}}{\prod_{i \in N}(i\,i{+}1)} \\
&\hspace{3cm}\rightarrow \int \frac{d^{2\times n}\sigma}{GL(2)} \diracd{2\times n}{SE^n_{L'}}\frac{\prod_{l\in L'} \diracd{\cN}{\eta_l - \sum_{r\in R'}\frac{\eta_r}{(lr)}}}{\prod_{i \in N}(i\,i{+}1)}(l_0r_0)^{\cN-4}. 
\end{split}
\end{equation}
For $\cN = 4$ the integrand is unchanged up to a choice of the left set, and hence for $\cN = 4$ super Yang-Mills theory we can see that the choice of left set does not affect the overall superamplitude. Now look at the MHV sector for $\cN = 0$, fix $L = \{1,2\}$ and swap the external data for particles 1 and $r_0$, as well as performing the integral transformation to swap 1 and $r_0$. We see that $(1r_0)^4 = \frac{\ang{12}^4}{\ang{2r_0}^4}$, which is exactly the factor required to modify the Park-Taylor formula for particles 1 and 2 negative helicity gluons to have particles 2 and $r_0$ with negative helicities. It is interesting to note that this structure extends outside of the MHV sector at the level of the integrand in the scattering equation formalism.

\subsection{The Jacobian of the Scattering Equations}
\label{sec:jac}

In this appendix we detail some properties of the Jacobian of the 4D scattering equations. For a general worldsheet integral over the scattering equation delta functions, we can solve the integrals and write the expression as a sum over the solutions to the scattering equations. To do this we need an explicit expression for the Jacobian $J^n_L(\sigma)$ as follows. We use the notation for matrix $A$ that $A^{ij}$ has rows and columns $i$ and $j$ removed. Let~$l,l'\in L$ and delete equations $l$ and $l'$ to arrive at

\begin{equation}
\int \frac{d^{2\times n}\sigma}{GL(2)} \diracd{2\times n}{SE^n_L} f(\sigma) = \delta^4(P) \sum_{\sigma_{\rm sol} \in {\rm solutions}} \frac{ f(\sigma_{\rm sol})}{\ang{ll'}^{-2}\det(J^{n\,\,ll'}_{L}(\sigma_{\rm sol}))}.
\end{equation}

Note that as shown in \aref{sec:gfix}, the two rows/columns deleted must either both be in the left set or both in the right set, and there is an extra associated factor eg. $\ang{ll'}^2$ such that the full determinant of the Jacobian is $\ang{ll'}^{-2}\det(J^{n\,\,ll'}_{L}) $ for two left set particles deleted and gauge-fixed.

At $n$ points with left set $L$ we find the Jacobian to be

\begin{equation}
J^n_L = \begin{pmatrix}\pdd{\tilde{E}_l}{\sigma_{l'}}&\pdd{\tilde{E}_l}{\sigma_{r'}}\\
\pdd{E_r}{\sigma_{l'}}&\pdd{E_r}{\sigma_{r'}}\end{pmatrix} = 
 \begin{pmatrix}-\delta_{ll'}\sum_{r\in R}\frac{\sket{r}\otimes\sigma_r}{(lr)^2} &\frac{\sket{r'}\otimes \sigma_l}{(lr')^2}\\
\frac{\bra{l}\otimes \sigma_r}{(lr)^2}&-\delta_{rr'}\sum_{l\in L}\frac{\bra{l}\otimes \sigma_l}{(lr)^2}\end{pmatrix},
\end{equation}

\noindent where the matrix is written in a block form with blocks of sizes of the left and right set, and each element of these matrices is broken down into a $2\times 2$ matrix which is a tensor product of a spinor with a worldsheet vector. This matrix has determinant 0, which is simple to check analytically for example in {\sc Mathematica}. This is insured by the fact that gauge-fixing removes 4 of the sigma variables, and hence we have to remove four of the rows of the matrix to produce a well specified system. Similarly we must remove four of the columns of $J^n_L$, which is equivalent to deleting two equations as shown in \aref{sec:gfix}.

In the MHV sector this matrix is block diagonal and we can calculate the determinant in terms worldsheet minors directly. Taking the left set to the particles 1 and 2 and also gauge-fixing and deleting these particles we find that

\begin{equation}
\det\left(J^{n\hspace{.3cm}12}_{\{1,2\}}\right) = \det\left( \pdd{E_r}{\sigma_{r'}} \right) = \prod_{r\in R}\det\left( \frac{\bra{1}\otimes\sigma_1}{(1r)^2} + \frac{\bra{2}\otimes\sigma_2}{(2r)^2}\right).
\end{equation}
We now use the general result for determinants of sums of tensor products of two dimensional vectors that
\begin{equation}
\label{eq:2by2matrixdet}
\det\left(\sum_{i = 1}^m \alpha_i u_i \otimes v_i \right) = \sum_{1 \leq i < j \leq m}\alpha_i \alpha_j \det(u_i u_j) \det(v_i v_j),
\end{equation}

\noindent for $m$ variables $u_i, v_i \in \mathbb{C}^2$ and $\alpha_i \in \mathbb{C}$. For the MHV calculation $m=2$, giving

\begin{equation}
\det\left(J^{n\hspace{.3cm}12}_{\{1,2\}}\right) = \prod_{r\in R}\frac{\ang{12}(12)}{(1r)^2(2r)^2} =  \prod_{r\in R}\frac{\ang{1r}^2\ang{2r}^2}{\ang{12}^3},
\end{equation}

\noindent where in the last equation we substitute in the MHV solution from \sref{sec:analytical}.

It is also possible to analytically evaluate the Jacobian outside of the MHV sector. Assume $1, 2 \in L$ and gauge-fix and delete particles 1 and 2, and use the formula for the determinant of a block matrix to find that 
\begin{equation}
\det\left(J^{n\,12}_{L}\right) = \det\left(\pdd{E_r}{\sigma_{r'}}\right)
\det\left(
\pdd{\tilde{E}_l}{\sigma_{l'}}- \pdd{\tilde{E}_l}{\sigma_{r'}}\left(\pdd{E_r}{\sigma_{r'}}\right)^{-1}\pdd{E_r}{\sigma_{l'}}
\right),
\end{equation}

\noindent where $r, r' \in R$ and $l, l' \in L/\{1,2\}$.

As $\pdd{E_r}{\sigma_{r'}}$ is block diagonal, it is comparatively simple to calculate its determinant and inverse. Using \eref{eq:2by2matrixdet} we find the determinant to be
\begin{equation}
 \det\left(\pdd{E_r}{\sigma_{r'}}\right) = \prod_{r\in R}\sum_{l<l' \in L}\frac{\ang{ll'}(ll')}{(rl)^2(rl')^2}.
\end{equation}

\noindent To invert this matrix we see that in the $r,r'$ indices it is simply $\delta_{rr'}$, leaving the calculation of the inverses of the $2\times2$ blocks. We use the result for a $2\times2$ matrix $M$ that $M^{-1} = \det(M)^{-1}\left(\begin{smallmatrix}0&1\\-1&0\end{smallmatrix}\right)M\left(\begin{smallmatrix}0&-1\\1&0\end{smallmatrix}\right)$ to invert in the spinor and worldsheet indices. This corresponds to raising and lowering the two indices, and dividing by the determinant as calculated by \eref{eq:2by2matrixdet}. The result is then that 

\begin{equation}
\begin{split}
&\det\left(
\pdd{\tilde{E}_l}{\sigma_{l'}}- \pdd{\tilde{E}_l}{\sigma_{r'}}\left(\pdd{E_r}{\sigma_{r'}}\right)^{-1}\pdd{E_r}{\sigma_{l'}}
\right)
\\& = 
\det\left(
\sum_{r\in R}\frac{\sbra{r}\otimes\sigma_r}{(lr)^2\sum_{\lambda<\lambda'\in L}\frac{\ang{\lambda\lambda'}(\lambda\lambda')}{(r\lambda)^2(r\lambda')^2}}
\left(\sum_{\lambda \in L}\frac{\ang{l'\lambda}(l \lambda)}{(\lambda r)^2(l'r)^2} -
\delta_{ll'}\sum_{\lambda<\lambda'\in L}\frac{\ang{\lambda\lambda'}(\lambda\lambda')}{(\lambda r)^2(\lambda'r
)^2}
\right)\right),
 \end{split}
\end{equation}
where the determinant is taken over $l,l' \in L/\{1,2\}$, combined with the tensor product of spinor and worldsheet indices. In the NMHV sector we see that $l = l' = 3$ and we can calculate the determinant of the remaining $2\times2$ matrix using \eref{eq:2by2matrixdet}. As an example at 6 points NMHV we find that

\begin{equation}
\det(J^{6\hspace{.5cm}12}_{\{1,2,3\}}) = \frac{\ang{12}^2(12)^2}{\prod_{l,r}(lr)^2}\sum_{r<r'\in R\atop l<l' \in L}\squ{rr'}(rr')\ang{ll'}(ll')\sum_{\lambda\in L\atop\rho\in R}\epsilon_{ll'\lambda}\epsilon_{rr'\rho}(\lambda\rho)^2.
\end{equation}

\bibliographystyle{ieeetr} 
\bibliography{4dscatteqnsreferences}
\end{document}